\documentclass[journal=nalefd,manuscript=letter,email=True,layout=twocolumn]{achemso}

\usepackage{graphicx}
\usepackage{latexsym}
\usepackage{amsmath}
\usepackage{amssymb}
\usepackage{amsfonts}
\usepackage{color}
\usepackage{bm}
\usepackage{verbatim}
\usepackage{pagecolor,lipsum}
\usepackage{float}
\usepackage{hyperref}
\usepackage[normalem]{ulem}
\hypersetup{colorlinks=true,urlcolor=blue,linkcolor=blue,citecolor=blue}
\usepackage{comment}
\usepackage{soul,xcolor}

\usepackage{xcolor}

\definecolor{MS-color}{RGB}{128,0,128}

\setstcolor{red}

\usepackage{graphicx}
\usepackage{latexsym}
\usepackage{amsmath}
\usepackage{amssymb}
\usepackage{amsfonts}
\usepackage{color}
\usepackage{bm}
\usepackage{verbatim}
\usepackage{pagecolor,lipsum}
\usepackage{float}
\usepackage{hyperref}
\usepackage[normalem]{ulem}
\hypersetup{colorlinks=true,urlcolor=black,linkcolor=black,citecolor=black}
\usepackage{comment}

\author{Remko Fermin}
\author{Dyon van Dinter}
\author{Michel Hubert}
\author{Bart Woltjes}
\affiliation{Huygens-Kamerlingh Onnes Laboratory, Leiden University, P.O. Box 9504, 2300 RA Leiden, The Netherlands.}
\author{Mikhail Silaev}
\affiliation{Department of Physics and Nanoscience Center, University of Jyv\"askyl\"a, P.O.
Box 35 (YFL), Jyv\"askyl\"a, Finland}
\alsoaffiliation{Computational Physics Laboratory, Physics Unit, Faculty of Engineering and Natural Sciences, Tampere University, P.O. Box 692, Tampere, Finland}
\author{Jan Aarts}
\author{Kaveh Lahabi}
\affiliation{Huygens-Kamerlingh Onnes Laboratory, Leiden University, P.O. Box 9504, 2300 RA Leiden, The Netherlands.}
\email{lahabi@physics.leidenuniv.nl}

\title[Rim currents in a ferromagnetic disk]{Superconducting triplet rim currents in a spin-textured ferromagnetic disk}

\begin{document}


\begin{abstract}
Since the discovery of the long-range superconducting proximity effect, the interaction between spin-triplet Cooper pairs and magnetic structures such as domain walls and vortices has been the subject of intense theoretical discussions, while the relevant experiments remain scarce. We have developed nanostructured Josephson junctions with highly controllable spin texture, based on a disk-shaped Nb/Co bilayer. Here, the vortex magnetization of Co and the Cooper pairs of Nb conspire to induce long-range triplet (LRT) superconductivity in the ferromagnet. Surprisingly, the LRT correlations emerge in highly localized (sub-80 nm) channels at the rim of the ferromagnet, despite its trivial band structure. We show that these robust rim currents arise from the magnetization texture acting as an effective spin-orbit coupling, which results in spin accumulation at the bilayer-vacuum boundary. Lastly, we demonstrate that by altering the spin texture of a single ferromagnet, both $0$ and $\pi$-channels can be realized in the same device.
\end{abstract}


\section{Introduction}

The appearance of localized supercurrents at the edges of a Josephson junction is generally attributed to the topology of the electronic band structure and edge states.\cite{Sato2017}. Edge states and the accompanying edge currents are typically found in ultraclean systems such as 2D electron gases,\cite{Hart2014} nanowires,\cite{Murani2017} and graphene.\cite{Allen2016} Here, we report the emergence of highly localized (sub-80~nm) spin-polarized supercurrents at the rim of disk-shaped Josephson junctions with a diffusive ferromagnetic barrier (Co). As we demonstrate, however, the rim currents are not related to the electronic band structure; but rather a direct result of the interactions between spin-triplet Cooper pairs and the nontrivial spin texture of the ferromagnet.

At the interface between a superconductor and a ferromagnet, short-range triplet (SRT) Cooper pairs with zero spin projection emerge naturally via spin-mixing of singlet pairs, and decay over a few nanometers ($\xi_{\text{F}} (\text{Co}) \sim 3$ nm \cite{Robinson2006}) inside the ferromagnet. Long-range triplet (LRT) pairs can, on the other hand, propagate over substantially larger distances.\cite{Bergeret2001a,Bergeret2005} Half-metallic systems can even show the LRT proximity effect over hundreds of nanometers.\cite{Eschrig2008,Keizer2006,Anwar2010,Singh2016} Due to their spin polarization, the LRT Cooper pairs can provide the means to combine the absence of Joule heating and decoherence with the functionality of spintronic devices.\cite{Linder2015,Eschrig2015a} However, the controlled generation of LRT currents has proven to be a demanding process, commonly realized in complex superconductor-ferromagnet (S$-$F) hybrids, involving multiple F layers with noncollinear magnetization.\cite{Houzet2007,Khaire2010,Robinson2010,Anwar2012,Leksin2012,Iovan2014,Martinez2016,Lahabi2017a,Kapran2020,Aguilar2020,Komori2021}. Furthermore, a substantial body of research considered the possibility of generating and controlling LRT correlations using spin-textured systems, such as domain walls\cite{Bergeret2001a,Fominov2007,Volkov2008,Kalcheim2011,Aikebaier2019} and vortices.\cite{Silaev2009,Kalenkov2011} However, the experimental evidence to verify such models remains scarce.\cite{Bhatia2021}
In other recent developments, it was suggested that theoretically spin mixing can also be achieved by spin-orbit coupling (SOC)\cite{Niu2012,Bergeret2013_new,Bergeret2014a,Alidoust2015_discus1,Jacobsen2015}. This led to researching long-range proximity effects with Josephson junctions containing heavy  metal interlayers.\cite{Satchell2018,Jeon2018,Bujnowski2019,Eskilt2019,Jeon2020,Satchell2019,Silaev2020}. In addition, recent studies suggest that spin-orbit coupling (SOC) can lead to spin accumulation at the edges of Josephson devices\cite{Alidoust2015_discus1,Tokatly2019,bobkova2004effects} and, in some cases, generation of LRT currents.\cite{Alidoust2015_discus2,Salamone2021,Mazanik2022,Bobkova2021_new} At present there is a complete lack of experiments that can examine the influence of SOC on LRT transport. As a consequence, the interplay between triplet pairing and magnetic texture as well as SOC remains elusive.

 \begin{figure}[b!]
 \centerline{$
 \begin{array}{c}
  \includegraphics[width=0.8\linewidth]{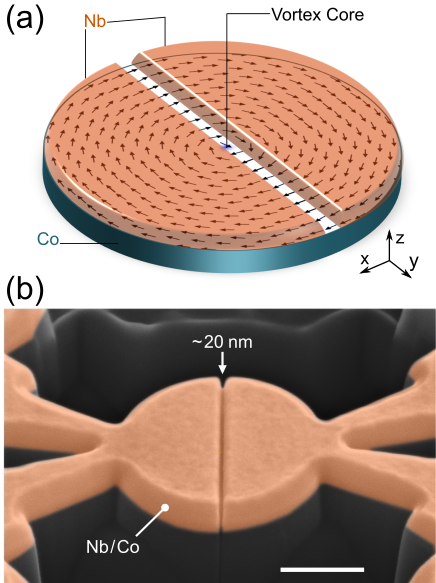}
 \end{array}$}
 \caption{\textbf{a.} Schematic of the Josephson device. The Nb electrodes are separated by a trench, forming a Co weak link. The pattern on the Co layer corresponds to micromagnetic simulations of a micrometer-size disk. (for more information on the micromagnetic simulations, see Supporting Information sec. S2). The arrows correspond to the in-plane magnetization, while the out-of-plane component is represented by color, which only appears at the vortex core (blue region; less than 5 nm in diameter). \textbf{b.} False colored scanning electron micrograph of a structured bilayer. The 20 nm gap indicates the Co weak link at the bottom of the trench. The scale bar is equivalent to 400 nm.} \label{Co-fig1}
 \end{figure}  

\begin{figure*}[t!]
 \centerline{$
 \begin{array}{c}
  \includegraphics[width=1\linewidth]{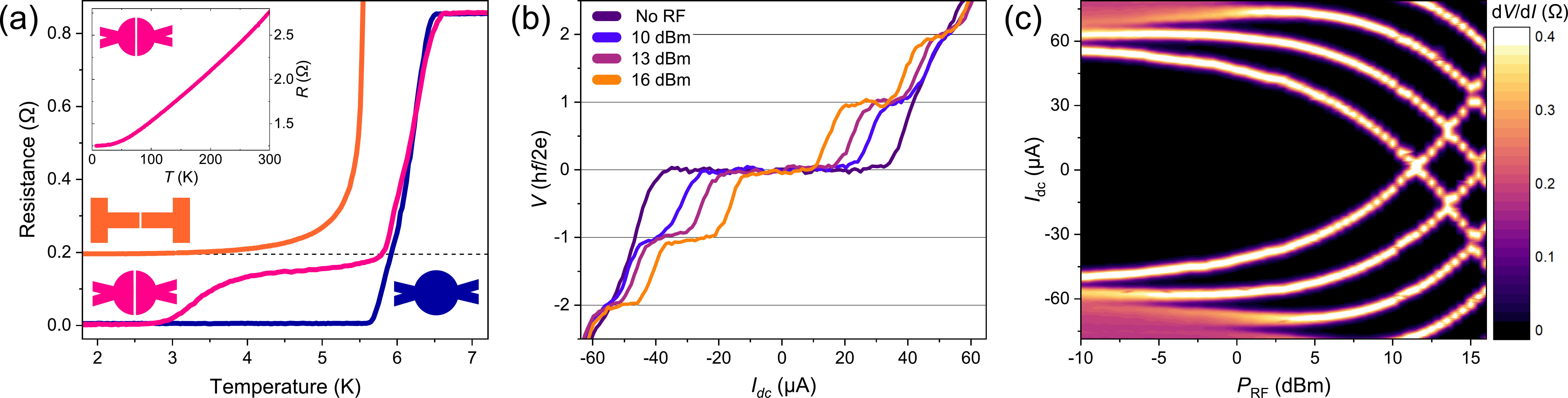}
  \end{array}$}
 \caption{Overview of the basic transport properties of the devices. \textbf{a.} Resistance as a function of temperature of a proximized disk device (pink) is compared to a Co-Nb disk without a trench (blue) and a bar-shaped device with uniform magnetization (orange). The bar-shaped device is not proximized due to the lack of spin texture (for more information, see Supporting Information sec. S3). \textbf{b.} $IV$-characteristics at 2 K of a proximized disk device, measured while irradiating the sample with $f=$ 1.1 GHz of RF radiation for different powers. The voltage is normalized in units of $hf/2e$. \textbf{c.} Evolution of the Shapiro response as a function of RF power represented as $\text{d}V/\text{d}I$ colormap.} \label{Co-fig2}
 \end{figure*} 

To address these issues, we have developed a disk-shaped S$-$F$-$S Josephson junction with a highly controllable ferromagnetic vortex spin texture, capable of converting singlet Cooper pairs into LRT currents (see Figure~\ref{Co-fig1}). The device consists of a Nb/Co bilayer, where a trench in the Nb layer introduces a ($\sim 20$ nm) cobalt weak link, which eliminates any singlet or SRT transport. We show that a magnetization gradient of the vortex can act as an effective SOC, which leads to spin accumulation at the rims of the device. This is verified by our transport experiments, which show that the LRT transport is highly localized at the rims of the ferromagnet, resulting in a distinct double-slit supercurrent interference pattern. By modifying the spin texture in a controllable manner, we show that both $0$ and $\pi$ segments can emerge in a single junction. Utilizing the linearized Usadel equation, we examine the microscopic origin of the rim currents in the proximized ferromagnet. Our findings suggest that, in addition to spin texture, superconductor-vacuum boundary conditions play an important role in the singlet to LRT conversion.

\section{Results and Discussion} 

\subsection{Establishing long-range triplet transport}

Figure~\ref{Co-fig2}a shows resistance as a function of temperature for a typical disk junction (see also Supporting Information sec. S1). A micrometer-wide weak link has a typical resistance of 200 m$\Omega$ and becomes fully proximized at low temperatures. We unambiguously establish the Josephson transport in our device by observing their Shapiro response to microwave radiation. This is carried out by measuring the current-voltage ($IV$)-characteristics while irradiating the junction with microwaves from a nearby antenna. The $IV$-curves show clear Shapiro steps (discrete voltage steps of $hf/(2e)$, where $h$ is the Planck constant, $f$ is the frequency, and $e$ is the electron charge), which is a result of the phase-locking between the applied microwaves and the Josephson currents (see figure~\ref{Co-fig2}b). We also examined the evolution of the width of the voltage plateaus as a function of microwave power. The results are presented in Figure~\ref{Co-fig2}c as a color map of differential resistance.

A direct method to examine the presence of LRT correlations is to verify that, once the mechanism for the emergence of LRT pairing is eliminated, the proximity effect will disappear. For instance, in the case of S/F’/F/F’/S multilayer junctions used in previous studies, where the generation of LRT correlations requires a magnetic noncollinearity between the F and F’ layers, the control experiment would show that the critical current ($I_\text{c}$) is heavily suppressed if the F’ layers were either removed or magnetized parallel (or antiparallel) to the F layer.\cite{Khaire2010,Martinez2016,Leksin2012,Glick2018,Aguilar2020,Robinson2010}
The same argument applies here: if the proximity effect is due to LRT correlations produced by the spin texture of the junction, the $I_\text{c}$ must vanish once the magnetization is uniform. This actually happens when we remove the spin texture by applying an in-plane field, typically around 100 mT, which completely suppresses the $I_\text{c}$ (see Supporting Figure S3a). We also verify this through two further experiments, described in Supporting Information sec. S3. First, we examined the transport in bar-shaped control samples, where shape anisotropy ensures that (even in the absence of in-plane fields) the cobalt layer has a uniform magnetization along the long axis of the bar (see Figure S1). These samples were fabricated via the same procedure as the primary disk-shaped junctions and received the same FIB treatment to structure their weak link. This is evident by the fact that the bar-shaped junctions and the disk devices have a matching barrier resistance ($\approx$ 200 m$\Omega$). Despite multiple attempts, however, the bar-shaped control samples show no sign of long-range proximity. Additionally, we also prepared disk-shaped control junctions where, by applying a lower dose of Ga ions when structuring the weak link, we leave some residual Nb at the bottom of the trench – forming a nonmagnetic channel for singlet transport. Such junctions are completely insensitive to the magnetic state of the cobalt disk and are robust against the in-plane fields used for altering the spin texture; they maintain their $I_\text{c}$ at fields as high as 2 T (see Supporting Figure S3b).

\subsection{Triplet currents confined to the rims of the disk}

We establish the presence of rim currents using superconducting quantum interferometry (SQI), \textit{i.e.}, measuring the critical current as a function of a magnetic field, applied perpendicular to the transport direction (out-of-plane). Note that the out-of-plane fields used in our SQI experiments are too small to disturb the stable vortex magnetization of the Co disk.

 \begin{figure*}[t!]
 \centerline{$
 \begin{array}{c}
  \includegraphics[width=1\linewidth]{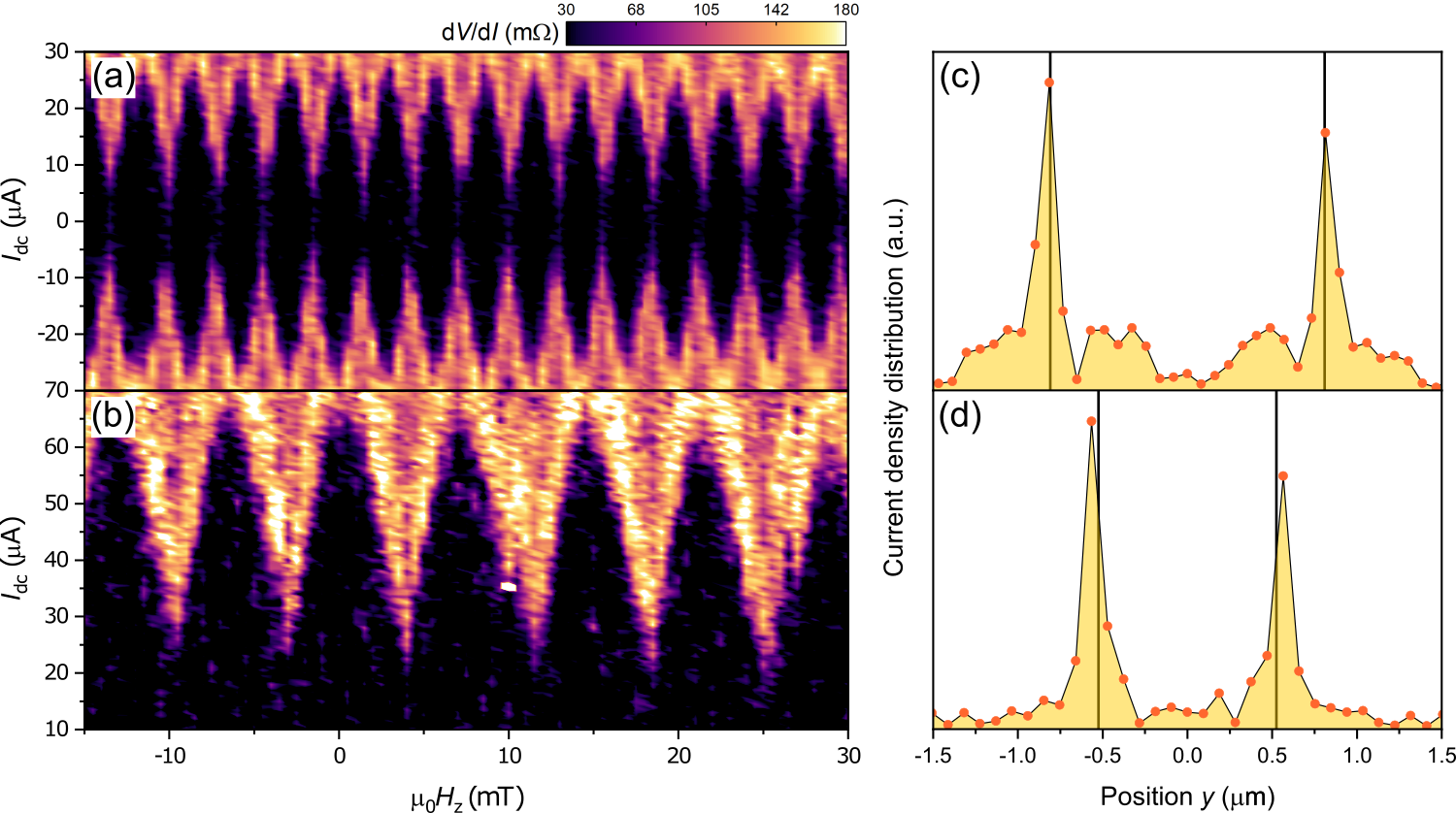}
 \end{array}$}
 \caption{\label{Fig:3} \textbf{a.} and \textbf{b.} The superconducting interference patterns of two junctions with different diameters. The pattern is obtained by measuring the differential resistance as a function of d.c. current and magnetic field. The disk diameter in \textbf{a.} and \textbf{b.} are 1.62 $\mu$m and 1.05 $\mu$m, respectively. The period of the oscillations scales inversely with the junction area. In both cases, the junctions show a clear two-channel interference pattern. \textbf{c.} and \textbf{d.} depict the critical current density profiles obtained by the Fourier analysis of the patterns in \textbf{a.} and \textbf{b.} respectively. The vertical lines indicate the boundaries of the device.} \label{Co-fig3}
 \end{figure*}   

In a conventional junction, the supercurrent is distributed uniformly across the weak link. This results in the well-known Fraunhofer SQI pattern, where the oscillation amplitude has a $1/B$ dependence, and the center lobe is twice as wide as its neighbors. As shown in Figure~\ref{Co-fig3}, our devices show a completely different behavior: two-channel interference patterns, characterized by equal-width lobes and slow decay of oscillation amplitude.

All the triplet junctions we measured (over ten devices) show such a two-channel interference pattern. This is illustrated in Figure~\ref{Co-fig3}, where we show the SQI patterns for two junctions with different diameters (1.62 $\mu$m and 1.05 $\mu$m). Note that the period of the oscillations scales inversely with the area of the junction, which is determined by the radius of the disk. We apply inverse Fourier transform to the SQI patterns to reconstruct the spatial distribution of supercurrent density.~\cite{Dynes1971} This is a well-established technique, commonly applied to verify the existence of edge currents (see also Supporting Information sec. S4).\cite{Suominen2017,Huang2019} Figure~\ref{Co-fig3} shows the results of our Fourier analysis for both devices. Regardless of the sample area, we consistently find the supercurrent to be highly localized at the rim of the sample (70 nm or less in width, limited by the resolution of the Fourier analysis). Furthermore, the channels are highly symmetric, as indicated by the sharp cusps of the SQI pattern. Note that the trench is deepest on the sides of the disk (due to a higher milling rate, as can be seen by the small notches on sides of the disk in Figure~\ref{Co-fig1}b), making the formation of accidental singlet edge channels even less probable. More importantly, the two-channel behavior is completely absent in all the singlet control samples (see Supporting Information Sec. S3). If the barrier contains residual Nb or is made out of a normal metal (silver), the junction yields a standard single-channel diffraction pattern (Supporting Figures S2 and S4b respectively).

 \begin{figure*}[hbt!]
 \centerline{$
 \begin{array}{c}
  \includegraphics[width=0.95\linewidth]{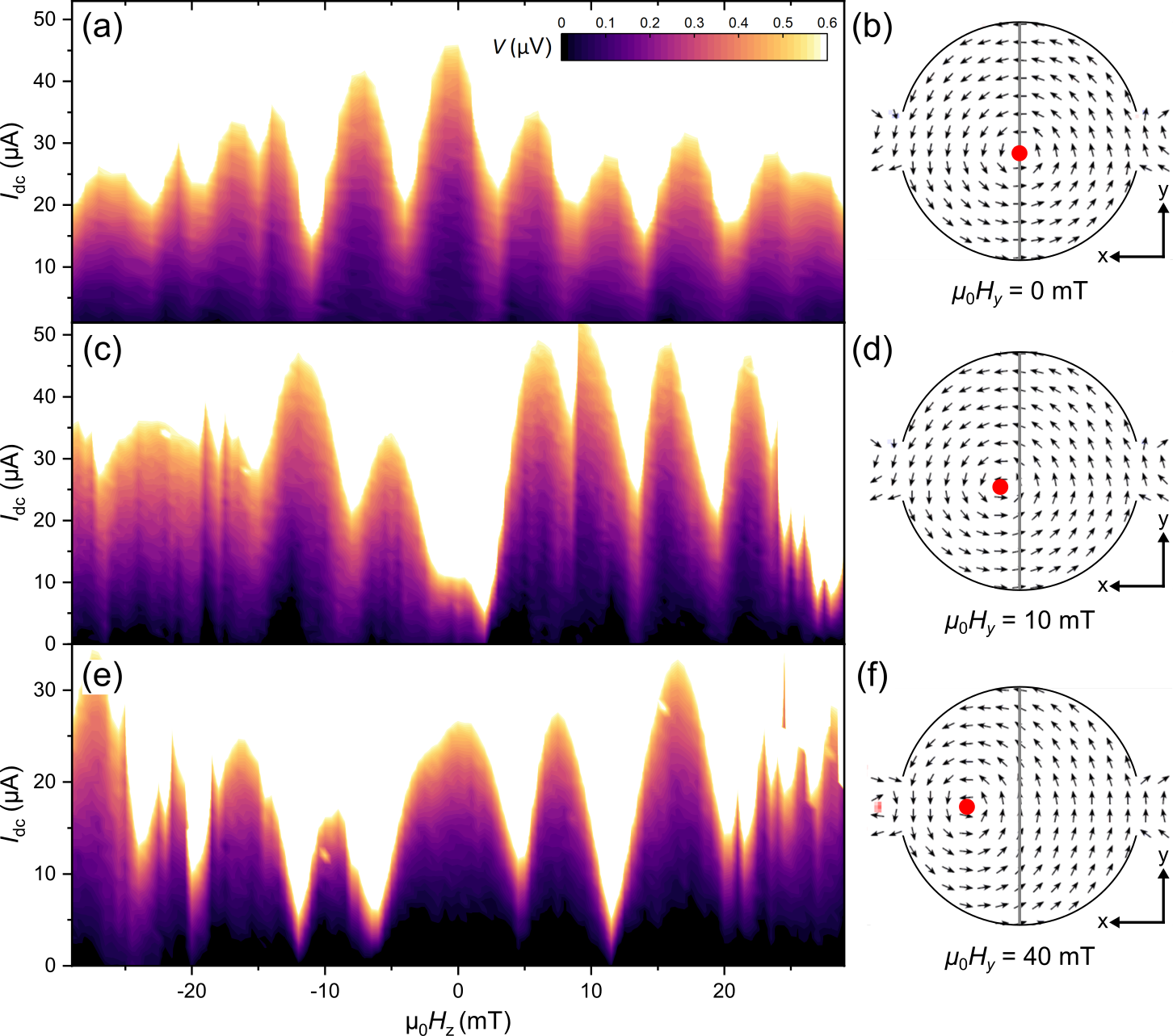}
 \end{array}$}
 \caption{\label{Fig:4} Supercurrent interference patterns (left column) measured at different in-plane fields and the corresponding simulated spin textures (right column). The gray line represents the position of the weak link, and the red dot indicates the location of the vortex core. \textbf{a.} and \textbf{b.} At zero in-plane field, the vortex is at the center of the disk, and a SQUID pattern is observed, \textit{i.e.}, lobes of equal width and slow decay of peak height. \textbf{c.} and \textbf{d.} Applying $\mu_0 H_\text{y}$ = 10 mT breaks the axial symmetry of the vortex magnetization. This results in the suppression of the middle peak in the interference pattern, characteristic of a $0-\pi$ SQUID. \textbf{e.} and \textbf{f.} At $\mu_0 H_\text{y}$ = 40 mT the vortex core is displaced by over 100 nm. The middle peak of the interference pattern is recovered and its width is doubled with respect to the original pattern. The peak height increases as a function of the out-of-plane field, regardless of sweep direction.} \label{Co-fig4}
 \end{figure*}   

\subsection{Altering the magnetic texture by an in-plane field}

So far we have shown the unconventional distribution of supercurrents through the spin-textured ferromagnetic weak link. We investigate this further by modifying the spin texture using an in-plane magnetic field. As shown in our micromagnetic simulations, in-plane fields can alter the spin texture by effectively moving the vortex core along the axis perpendicular to the field (see Figure~\ref{Co-fig4}). For small in-plane fields, the core displacement has an almost linear response and is fully reversible. Using a vector magnet system, we are able to apply a constant in-plane magnetic field while simultaneously acquiring the SQI pattern as described above. Figure \ref{Co-fig4} shows the SQI patterns measured for different in-plane fields, applied along the trench (y-axis), together with the corresponding micromagnetic simulations.

At zero in-plane field (Figure~\ref{Co-fig4}a), we observe the aforementioned two-channel (SQUID) interference pattern, with maximum $I_\text{c}$ at zero field. At $\mu_0 H_\text{y}$ = 10 mT we observe a strong suppression of $I_\text{c}$ for zero out of plane field. Remarkably, however, $I_\text{c}$ is recovered upon increasing the out-of-plane field in either direction. The resulting SQI pattern bears resemblance of a $0-\pi$ SQUID: all the lobes are similar in width and $I_\text{c}$ is suppressed around zero. Increasing the $H_\text{y}$ to 40 mT, the vortex core has traveled over 100 nm away from the center of the junction (Figure~\ref{Co-fig4}e). Interestingly, we find the $I_\text{c}$ to recover for zero out-of-plane field. This reentrant behavior is accompanied by drastic changes to the SQI pattern. The central lobe is now twice as wide, indicating a modified supercurrent distribution, which no longer corresponds to the original two channels. Even more striking is the amplitude of the $I_\text{c}$ oscillations: instead of decaying, the lobes grow taller as we increase the magnitude of the out-of-plane field. This is the universal characteristic of $0-\pi$ junctions, i.e., a junction consisting of multiple $0$ and $\pi$ segments connected in parallel.\cite{zigzag1,zigzag2,zigzag3} Note that the observed evolution of the interference pattern with the in-plane field cannot be attributed to stray fields or misalignment with the magnet axes since the SQI patterns are independent of magnetic field sweep direction. More importantly, the behavior is completely absent in the control samples with no LRT (Supporting Information Sec. S3), which yield Fraunhofer patterns, regardless of the amplitude or direction of the in-plane field.

\subsection{Mapping spin texture to spin-orbit coupling}

We now continue by describing the mechanism behind the formation of the rim currents using a model that links the vortex spin texture to the SOC. Within the cobalt weak link, the gradient of the local spin texture of the disk junctions is, at the rims of the device, not substantially larger than that of a bar-shaped device. This implies that even though the LRT currents emerge at the rims, they are formed by a process that is sensitive to the global spin texture of the disk.

It was demonstrated that the combination of SOC and exchange field \cite{bobkova2004effects,Alidoust2015_discus1,Salamone2021} or Zeeman field \cite{Tokatly2019,bergeret2020} can result in an equilibrium spin current (ESC) which accumulates at the superconducting/vacuum boundaries. We show that a similar process occurs in the presence of spin texture $\bm m (r)$, which produces the pure gauge SU(2) field that acts as effective spin-orbit coupling denoted by $   i  \hat U^\dagger \nabla \hat U $.\cite{tokatly2008equilibrium,Bergeret2013_new,Bergeret2014a} Here, $\hat U (\bm r)$ is the spin-rotation matrix, determined by the transformation to the local spin quantization axis. In the Supporting Information Section S5, where we present the full technical details of the Usadel calculations, we show that the vortex spin texture $\bm m = (-\sin\theta_v, \cos\theta_v, 0)$ can be transformed into a uniform one, with an effective SOC term $i  \hat U^\dagger \nabla \hat U = -\hat\sigma_z \nabla \theta_v/2$ (here $\hat\sigma_z$ is the spin Pauli matrix and $\theta_v = \arctan [(y-y_v)/(x-x_v)]$, where $x_v,y_v$ are the coordinates of the vortex center). Hence, our system is analogous to one with a uniform magnetization and an intrinsic, spatially inhomogeneous, SOC with the amplitude $|\nabla \theta_v|=1/r_v$  (with $r_v=\sqrt{(x-x_v)^2+(y-y_v)^2}$), and therefore hosts the aforementioned ESC.

The ESC is carried by the SRT Cooper pairs, which spontaneously appear both at the bottom of the S electrodes and the top of the F layer. The ESC can be thought of as a spin-imbalance in this SRT condensate. It can be parametrized in terms of the spin vector $\bm f$, which characterizes the spin component of the triplet condensate. The SRT pairing corresponds to $\bm f =  \bm m f_{\text{SRT}}$, while the LRT one is described by $\bm f_{\text{LRT}} \perp \bm m$. The direction of the ESC is determined by the in-plane gradients of the magnetic texture and flows parallel to the S$-$F interface. In terms of $\bm f$ it then becomes $ J^\gamma_j \propto |f_{\text{SRT}}|^2 (\bm m \times \nabla_j \bm m)_\gamma$ (here $\gamma= x,y,z$ is  
 the index in spin space and $j=x,y,z $ is the index in coordinate space). A ferromagnetic vortex texture yields $J^z_j \propto |f_{\text{SRT}}|^2 \nabla_i \theta_v = |f_{\text{SRT}}|^2   m_j /r_v $ which is in accordance with the general gauge-invariant expression for the spin current.\cite{10.21468/SciPostPhys.10.3.078}

\subsection{Mechanism for generating LRT rim currents}

Having established the equivalency between spin texture and SOC, we now provide a possible mechanism that relates the ESC to the emergence of LRT rim currents (see Supporting Figure S7 for a schematic representation). In the absence of spin texture, there is no ESC (Figure S7a). If the spin texture gradient is nonzero, the ESC adiabatically follows the local spin gauge field ($ \bm J^z \parallel \nabla\theta_v = (\bm m\times \nabla \bm m)_z $ see Figure S7b). When the spin current encounters the bilayer-vacuum boundary (for instance, due to deviations from the ideal circular geometry) the adiabatic approximation breaks down, resulting in an accumulation of spin at the rims of the device. Naturally, the spin accumulation decays over the spin diffusion length, which for cobalt is approximately 60 nm.\cite{Bass_2007} Near the interface, the adiabatic ESC can develop a nonzero normal component $ \bm m \times ( n_j \nabla_j) \bm m \neq 0$, where $\bm n$ is the interface normal. Since the total spin current is zero across the boundary, $n_j  J_j^z =0$, the adiabatic approximation breaks down and the ESC is compensated by a spin current carried by an LRT condensate, which emerges near the vacuum boundary (Figure S7c). Indeed, considering the local spin basis with $\bm m \parallel \bm x$ the vector product of $\bm f_{\text{SRT}} \parallel \bm m$ and $\nabla_j\bm f_{\text{LRT}}\perp \bm m $  provides the contribution to the $z$-component of the  spin current (mediated by the condensate $J_j^z \sim (\bm m \times \nabla_j \bm m)_z $) due to the first term of Eq.14 in the Supporting Information sec. S5. This contribution compensates the ESC near the rim. 
A similar SRT to LRT conversion process has been proposed to occur at the sample boundaries of SNS junctions with intrinsic SOC and a spin active interface\cite{Alidoust2015_discus2} or in one-dimensional systems with a geometric curvature.\cite{Salamone2021}

\begin{figure*}[t!]
 \centerline{$
 \begin{array}{c}
  \includegraphics[width=1\linewidth]{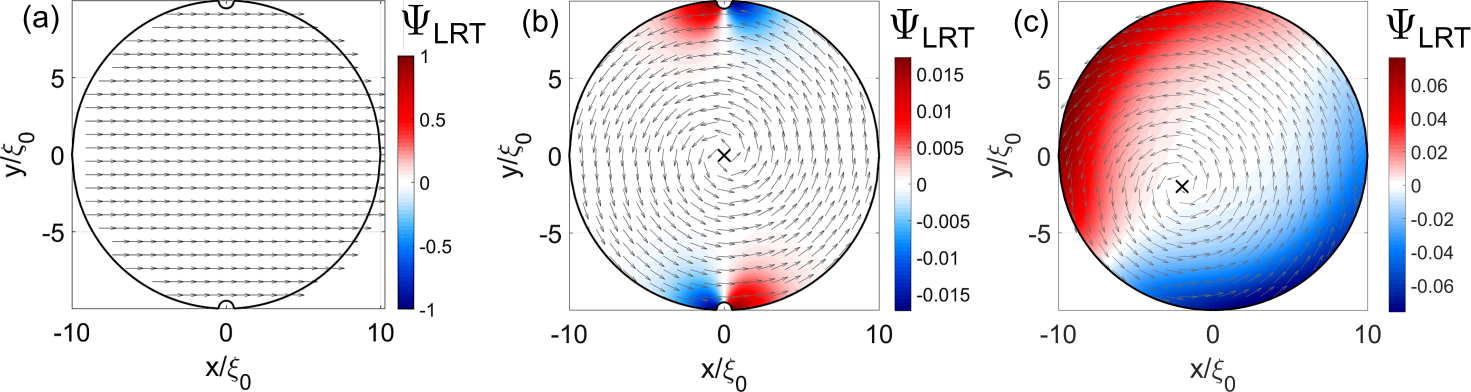}
 \end{array}$}
 \caption{\label{Fig:5} Simulated distribution of $\Psi_{\text{LRT}}$ in a S/F bilayer for three different spin configurations, where the amplitude is normalized to $ (\gamma \xi_0) f_{\text{SRT}}  $ ($\gamma$ is the S$-$F interface transparency). The direction of the in-plane spin texture is shown by arrows. The simulated geometry includes notches that arise from FIB processing of the actual devices. \textbf{a.} A homogeneous magnetization, where LRT correlations are completely absent. \textbf{b.} In the presence of a magnetic vortex, LRT Cooper pairs are generated at the notches, forming two LRT current channels. \textbf{c.} If the vortex core is shifted from the disk center, an asymmetry arises between the signs of the LRT channels, resulting in $0-\pi$ SQUID-like configuration, also in the absence of the notches.}
 \end{figure*}

By solving the linearized Usadel equation for a 2D disk-shaped S$-$F bilayer (without the trench), we simulate the distribution of the LRT amplitude $\Psi_{\text{LRT}}$, where $\bm f_{\text{LRT}} = \Psi_{\text{LRT}} (m_y, -m_x,0) $. The results for three different spin textures are presented in Figure \ref{Fig:5}. For a uniform magnetization, the LRT correlations are completely absent, regardless of sample geometry (Figure~\ref{Fig:5}a).
For a perfectly symmetric vortex pattern, any deviation from the ideal circular geometry at the sample-vacuum boundary (\textit{i.e.}, rim roughness or disorder) results in the emergence of LRT correlations. In our simulation (Figure~\ref{Fig:5}a,b), we use notches on the sides of the disk, also present in our devices, to demonstrate the effect of nonideal boundaries. However, in practice, any deviation from the perfect circular geometry or disorder at boundaries results in a similar outcome. Interestingly, even in the ideal circular geometry with flawless boundaries (\textit{i.e.}, atomically clean and smooth edges), the LRT currents would still appear if 

\noindent the magnetic vortex is not perfectly centered (see Figure~\ref{Fig:5}c). Note that for perfect circular symmetry, with the vortex at the center, $J^z_j$ will always remain parallel to $m_j$ and no LRT is generated.
 
Our simulations also provide insights into the phase of LRT correlations. When the vortex core is aligned with the trench (at $x=0$), the LRT currents will result in two $\pi$-channels, as indicated by the sign change of $\Psi_{\text{LRT}}$. In Figure \ref{Fig:5}c the vortex is displaced from the center (\textit{e.g.}, due to an in-plane field) the LRT channels develop opposite signs at the trench. This asymmetry is consistent with the observed $0-\pi$ SQUID interference pattern, measured under a constant in-plane field (Figure~\ref{Co-fig4}).

\subsection{Discussion}

While the model presented here can describe the emergence of LRT correlations at the bilayer-vacuum boundaries and the appearance of spontaneous supercurrents in our junctions (\textit{i.e.}, $0-\pi$ segments), we should point out that this formalism is restricted to 2D slices of the bilayer. Accounting for the superconductor-vacuum interface formed by the trench is more challenging, as it requires a full three-dimensional model and the knowledge of the exact trench dimensions (\textit{e.g.}, its extent in the Co layer). We discuss this further in Supporting Information sec. S5.

It should be noted that there is a fundamental difference between the devices presented here and those reported in a previous work, where the disk-shaped junctions consisted of a magnetic multilayer (S$-$F$'-$F$-$F$''-$S).\cite{Lahabi2017a} In contrast to the Nb/Ni/Cu/Co/Cu/Ni/Nb junctions, where long-range proximity was the result of the magnetic noncollinearity between the Co and Ni layers, here the LRT correlations are generated directly by the spin texture of a single ferromagnet. This is evident by the fact that the multilayer devices were highly sensitive to magnetic conditioning of the Ni (1.5~nm) layer (\textit{e.g.}, the $I_\text{c}$ was irreversibly enhanced when the sample was conditioned) whereas in the case of disks with a single ferromagnet, the transport characteristics are unaltered by magnetic conditioning, since the vortex magnetization is the global ground state of the Co-disk; regardless of the magnetic history, the disk will revert to the vortex magnetization at zero field. This is also confirmed by our micromagnetic simulations. Furthermore, devices with and without the nickel layer exhibit radically different behavior as a function of in-plane fields. However, there are similarities: both devices show a double slit interference pattern, although the current channels are considerably more confined in the case of a single ferromagnetic weak link.

\section{Conclusions}

In summary, we have revealed an unexpected interplay between triplet superconductivity and magnetic texture, which manifests itself as LRT supercurrents localized at the rim of the ferromagnet. We elucidate the origin of the rim currents by mapping the magnetic texture to an effective SOC, which leads to the emergence of equilibrium spin currents – carried by the triplet Cooper pairs present at the S$-$F interface. We also propose a mechanism for converting the spin currents into equal-spin LRT correlations based on the breakdown of the adiabatic approximation at the sample-vacuum boundaries. Lastly, we show that the nature of LRT transport undergoes drastic changes when the spin texture is modified. As illustrated here, by application of relatively small magnetic fields, the same Josephson junction can be tuned to function as both standard ($0-0$) and $0-\pi$ SQUIDs. The capacity to control supercurrents with the spin texture of a single ferromagnetic layer opens exciting prospects for regulating transport in superconducting devices. Ferromagnetic vortices, in particular, can be manipulated by microwave frequencies in a controllable manner, making them a promising candidate for ultrafast dissipationless electronics.

\begin{acknowledgement}

This work was supported by the project "Spin texture Josephson junctions" (Project Number 680-91-128) and by the Frontiers of Nanoscience (NanoFront) program, which are both (partly) financed by the Dutch Research Council (NWO). It was also supported by EU Cost Action CA16218 (NANOCOHYBRI) and benefited from access to The Netherlands Centre for Electron Nanoscopy (NeCEN) at Leiden University.

\end{acknowledgement}


\begin{mcitethebibliography}{71}
\providecommand*\natexlab[1]{#1}
\providecommand*\mciteSetBstSublistMode[1]{}
\providecommand*\mciteSetBstMaxWidthForm[2]{}
\providecommand*\mciteBstWouldAddEndPuncttrue
  {\def\EndOfBibitem{\unskip.}}
\providecommand*\mciteBstWouldAddEndPunctfalse
  {\let\EndOfBibitem\relax}
\providecommand*\mciteSetBstMidEndSepPunct[3]{}
\providecommand*\mciteSetBstSublistLabelBeginEnd[3]{}
\providecommand*\EndOfBibitem{}
\mciteSetBstSublistMode{f}
\mciteSetBstMaxWidthForm{subitem}{(\alph{mcitesubitemcount})}
\mciteSetBstSublistLabelBeginEnd
  {\mcitemaxwidthsubitemform\space}
  {\relax}
  {\relax}

%
\bibitem[Sato and Ando(2017)Sato, and Ando]{Sato2017}
Sato,~M.; Ando,~Y. {Topological superconductors: a review}. \emph{Rep. Prog.
  Phys.} \textbf{2017}, \emph{80}, 076501\relax
\mciteBstWouldAddEndPuncttrue
\mciteSetBstMidEndSepPunct{\mcitedefaultmidpunct}
{\mcitedefaultendpunct}{\mcitedefaultseppunct}\relax
\EndOfBibitem
%
\bibitem[Hart \latin{et~al.}(2014)Hart, Ren, Wagner, Leubner, M{\"{u}}hlbauer,
  Br{\"{u}}ne, Buhmann, Molenkamp, and Yacoby]{Hart2014}
Hart,~S.; Ren,~H.; Wagner,~T.; Leubner,~P.; M{\"{u}}hlbauer,~M.;
  Br{\"{u}}ne,~C.; Buhmann,~H.; Molenkamp,~L.~W.; Yacoby,~A. {Induced
  superconductivity in the quantum spin Hall edge}. \emph{Nat. Phys.}
  \textbf{2014}, \emph{10}, 638\relax
\mciteBstWouldAddEndPuncttrue
\mciteSetBstMidEndSepPunct{\mcitedefaultmidpunct}
{\mcitedefaultendpunct}{\mcitedefaultseppunct}\relax
\EndOfBibitem
\bibitem[Murani \latin{et~al.}(2017)Murani, Kasumov, Sengupta, Kasumov, Volkov,
  Khodos, Brisset, Delagrange, Chepelianskii, Deblock, Bouchiat, and
  Gu{\'{e}}ron]{Murani2017}
Murani,~A.; Kasumov,~A.; Sengupta,~S.; Kasumov,~Y.~A.; Volkov,~V.~T.;
  Khodos,~I.~I.; Brisset,~F.; Delagrange,~R.; Chepelianskii,~A.; Deblock,~R.;
  Bouchiat,~H.; Gu{\'{e}}ron,~S. {Ballistic edge states in Bismuth nanowires
  revealed by SQUID interferometry}. \emph{Nat. Commun.} \textbf{2017},
  \emph{8}, 15941\relax
\mciteBstWouldAddEndPuncttrue
\mciteSetBstMidEndSepPunct{\mcitedefaultmidpunct}
{\mcitedefaultendpunct}{\mcitedefaultseppunct}\relax
\EndOfBibitem
%
\bibitem[Allen \latin{et~al.}(2016)Allen, Shtanko, Fulga, Akhmerov, Watanabe,
  Taniguchi, Jarillo-Herrero, Levitov, and Yacoby]{Allen2016}
Allen,~M.~T.; Shtanko,~O.; Fulga,~I.~C.; Akhmerov,~A.~R.; Watanabe,~K.;
  Taniguchi,~T.; Jarillo-Herrero,~P.; Levitov,~L.~S.; Yacoby,~A. {Spatially
  resolved edge currents and guided-wave electronic states in graphene}.
  \emph{Nat. Phys.} \textbf{2016}, \emph{12}, 128\relax
\mciteBstWouldAddEndPuncttrue
\mciteSetBstMidEndSepPunct{\mcitedefaultmidpunct}
{\mcitedefaultendpunct}{\mcitedefaultseppunct}\relax
\EndOfBibitem
%
\bibitem[Robinson \latin{et~al.}(2006)Robinson, Piano, Burnell, Bell, and
  Blamire]{Robinson2006}
Robinson,~J. W.~A.; Piano,~S.; Burnell,~G.; Bell,~C.; Blamire,~M.~G. {Critical
  Current Oscillations in Strong Ferromagnetic $\pi$ Junctions}. \emph{Phys.
  Rev. Lett.} \textbf{2006}, \emph{97}, 177003\relax
\mciteBstWouldAddEndPuncttrue
\mciteSetBstMidEndSepPunct{\mcitedefaultmidpunct}
{\mcitedefaultendpunct}{\mcitedefaultseppunct}\relax
\EndOfBibitem
%
\bibitem[Bergeret \latin{et~al.}(2001)Bergeret, Volkov, and
  Efetov]{Bergeret2001a}
Bergeret,~F.~S.; Volkov,~A.~F.; Efetov,~K.~B. Long-Range Proximity Effects in
  Superconductor-Ferromagnet Structures. \emph{Phys. Rev. Lett.} \textbf{2001},
  \emph{86}, 4096\relax
\mciteBstWouldAddEndPuncttrue
\mciteSetBstMidEndSepPunct{\mcitedefaultmidpunct}
{\mcitedefaultendpunct}{\mcitedefaultseppunct}\relax
\EndOfBibitem
\bibitem[Bergeret \latin{et~al.}(2005)Bergeret, Volkov, and
  Efetov]{Bergeret2005}
Bergeret,~F.~S.; Volkov,~A.~F.; Efetov,~K.~B. Odd triplet superconductivity and
  related phenomena in superconductor-ferromagnet structures. \emph{Rev. Mod.
  Phys.} \textbf{2005}, \emph{77}, 1321\relax
\mciteBstWouldAddEndPuncttrue
\mciteSetBstMidEndSepPunct{\mcitedefaultmidpunct}
{\mcitedefaultendpunct}{\mcitedefaultseppunct}\relax
\EndOfBibitem
\bibitem[Singh \latin{et~al.}(2016)Singh, Jansen, Lahabi, and Aarts]{Singh2016}
Singh,~A.; Jansen,~C.; Lahabi,~K.; Aarts,~J. {High-Quality $\text{CrO}_{2}$
  Nanowires for Dissipation-less Spintronics}. \emph{Phys. Rev. X}
  \textbf{2016}, \emph{6}, 041012\relax
\mciteBstWouldAddEndPuncttrue
\mciteSetBstMidEndSepPunct{\mcitedefaultmidpunct}
{\mcitedefaultendpunct}{\mcitedefaultseppunct}\relax
\EndOfBibitem
\bibitem[Keizer \latin{et~al.}(2006)Keizer, Goennenwein, Klapwijk, Miao, Xiao,
  and Gupta]{Keizer2006}
Keizer,~R.~S.; Goennenwein,~S. T.~B.; Klapwijk,~T.~M.; Miao,~G.; Xiao,~G.;
  Gupta,~A. {A spin triplet supercurrent through the half-metallic ferromagnet
  CrO$_2$}. \emph{Nature} \textbf{2006}, \emph{439}, 825\relax
\mciteBstWouldAddEndPuncttrue
\mciteSetBstMidEndSepPunct{\mcitedefaultmidpunct}
{\mcitedefaultendpunct}{\mcitedefaultseppunct}\relax
\EndOfBibitem
\bibitem[Anwar \latin{et~al.}(2010)Anwar, Czeschka, Hesselberth, Porcu, and
  Aarts]{Anwar2010}
Anwar,~M.~S.; Czeschka,~F.; Hesselberth,~M.; Porcu,~M.; Aarts,~J. {Long-range
  supercurrents through half-metallic ferromagnetic CrO$_2$}. \emph{Phys. Rev.
  B} \textbf{2010}, \emph{82}, 100501\relax
\mciteBstWouldAddEndPuncttrue
\mciteSetBstMidEndSepPunct{\mcitedefaultmidpunct}
{\mcitedefaultendpunct}{\mcitedefaultseppunct}\relax
\EndOfBibitem
\bibitem[Eschrig and L{\"{o}}fwander(2008)Eschrig, and
  L{\"{o}}fwander]{Eschrig2008}
Eschrig,~M.; L{\"{o}}fwander,~T. {Triplet supercurrents in clean and disordered
  half-metallic ferromagnets}. \emph{Nat. Phys.} \textbf{2008}, \emph{4},
  138\relax
\mciteBstWouldAddEndPuncttrue
\mciteSetBstMidEndSepPunct{\mcitedefaultmidpunct}
{\mcitedefaultendpunct}{\mcitedefaultseppunct}\relax
\EndOfBibitem
\bibitem[Linder and Robinson(2015)Linder, and Robinson]{Linder2015}
Linder,~J.; Robinson,~J. W.~A. Superconducting spintronics. \emph{Nat Phys}
  \textbf{2015}, \emph{11}, 307--315\relax
\mciteBstWouldAddEndPuncttrue
\mciteSetBstMidEndSepPunct{\mcitedefaultmidpunct}
{\mcitedefaultendpunct}{\mcitedefaultseppunct}\relax
\EndOfBibitem
\bibitem[Eschrig(2015)]{Eschrig2015a}
Eschrig,~M. Spin-polarized supercurrents for spintronics: a review of current
  progress. \emph{Rep. Prog. Phys.} \textbf{2015}, \emph{78}, 104501\relax
\mciteBstWouldAddEndPuncttrue
\mciteSetBstMidEndSepPunct{\mcitedefaultmidpunct}
{\mcitedefaultendpunct}{\mcitedefaultseppunct}\relax
\EndOfBibitem
\bibitem[Houzet and Buzdin(2007)Houzet, and Buzdin]{Houzet2007}
Houzet,~M.; Buzdin,~A.~I. Long range triplet Josephson effect through a
  ferromagnetic trilayer. \emph{Phys. Rev. B} \textbf{2007}, \emph{76},
  060504\relax
\mciteBstWouldAddEndPuncttrue
\mciteSetBstMidEndSepPunct{\mcitedefaultmidpunct}
{\mcitedefaultendpunct}{\mcitedefaultseppunct}\relax
\EndOfBibitem
\bibitem[Lahabi \latin{et~al.}(2017)Lahabi, Amundsen, Ouassou, Beukers,
  Pleijster, Linder, Alkemade, and Aarts]{Lahabi2017a}
Lahabi,~K.; Amundsen,~M.; Ouassou,~J.~A.; Beukers,~E.; Pleijster,~M.;
  Linder,~J.; Alkemade,~P.; Aarts,~J. {Controlling supercurrents and their
  spatial distribution in ferromagnets}. \emph{Nat. Commun.} \textbf{2017},
  \emph{8}, 2056\relax
\mciteBstWouldAddEndPuncttrue
\mciteSetBstMidEndSepPunct{\mcitedefaultmidpunct}
{\mcitedefaultendpunct}{\mcitedefaultseppunct}\relax
\EndOfBibitem
\bibitem[Kapran \latin{et~al.}(2020)Kapran, Iovan, Golod, and
  Krasnov]{Kapran2020}
Kapran,~O.~M.; Iovan,~A.; Golod,~T.; Krasnov,~V.~M. {Observation of the
  dominant spin-triplet supercurrent in Josephson spin valves with strong Ni
  ferromagnets}. \emph{Phys. Rev. Res.} \textbf{2020}, \emph{2}, 013167\relax
\mciteBstWouldAddEndPuncttrue
\mciteSetBstMidEndSepPunct{\mcitedefaultmidpunct}
{\mcitedefaultendpunct}{\mcitedefaultseppunct}\relax
\EndOfBibitem
\bibitem[Khaire \latin{et~al.}(2010)Khaire, Khasawneh, Pratt, and
  Birge]{Khaire2010}
Khaire,~T.~S.; Khasawneh,~M.~A.; Pratt,~W.~P.; Birge,~N.~O. {Observation of
  spin-triplet superconductivity in Co-based josephson junctions}. \emph{Phys.
  Rev. Lett.} \textbf{2010}, \emph{104}, 137002\relax
\mciteBstWouldAddEndPuncttrue
\mciteSetBstMidEndSepPunct{\mcitedefaultmidpunct}
{\mcitedefaultendpunct}{\mcitedefaultseppunct}\relax
\EndOfBibitem
\bibitem[Martinez \latin{et~al.}(2016)Martinez, Pratt, and Birge]{Martinez2016}
Martinez,~W.~M.; Pratt,~W.~P.; Birge,~N.~O. {Amplitude Control of the
  Spin-Triplet Supercurrent in S /F /S Josephson Junctions}. \emph{Phys. Rev.
  Lett.} \textbf{2016}, \emph{116}, 077001\relax
\mciteBstWouldAddEndPuncttrue
\mciteSetBstMidEndSepPunct{\mcitedefaultmidpunct}
{\mcitedefaultendpunct}{\mcitedefaultseppunct}\relax
\EndOfBibitem
\bibitem[Leksin \latin{et~al.}(2012)Leksin, Garif'Yanov, Garifullin, Fominov,
  Schumann, Krupskaya, Kataev, Schmidt, and B{\"{u}}chner]{Leksin2012}
Leksin,~P.~V.; Garif'Yanov,~N.~N.; Garifullin,~I.~A.; Fominov,~Y.~V.;
  Schumann,~J.; Krupskaya,~Y.; Kataev,~V.; Schmidt,~O.~G.; B{\"{u}}chner,~B.
  {Evidence for triplet superconductivity in a superconductor-ferromagnet spin
  valve}. \emph{Phys. Rev. Lett.} \textbf{2012}, \emph{109}, 057005\relax
\mciteBstWouldAddEndPuncttrue
\mciteSetBstMidEndSepPunct{\mcitedefaultmidpunct}
{\mcitedefaultendpunct}{\mcitedefaultseppunct}\relax
\EndOfBibitem
\bibitem[Aguilar \latin{et~al.}(2020)Aguilar, Korucu, Glick, Loloee, Pratt, and
  Birge]{Aguilar2020}
Aguilar,~V.; Korucu,~D.; Glick,~J.~A.; Loloee,~R.; Pratt,~W.~P.; Birge,~N.~O.
  Spin-polarized triplet supercurrent in Josephson junctions with perpendicular
  ferromagnetic layers. \emph{Phys. Rev. B} \textbf{2020}, \emph{102},
  024518\relax
\mciteBstWouldAddEndPuncttrue
\mciteSetBstMidEndSepPunct{\mcitedefaultmidpunct}
{\mcitedefaultendpunct}{\mcitedefaultseppunct}\relax
\EndOfBibitem
\bibitem[Robinson \latin{et~al.}(2010)Robinson, Witt, and
  Blamire]{Robinson2010}
Robinson,~J. W.~A.; Witt,~J. D.~S.; Blamire,~M.~G. Controlled Injection of
  Spin-Triplet Supercurrents into a Strong Ferromagnet. \emph{Science}
  \textbf{2010}, \emph{329}, 59\relax
\mciteBstWouldAddEndPuncttrue
\mciteSetBstMidEndSepPunct{\mcitedefaultmidpunct}
{\mcitedefaultendpunct}{\mcitedefaultseppunct}\relax
\EndOfBibitem
\bibitem[Anwar \latin{et~al.}(2012)Anwar, Veldhorst, Brinkman, and
  Aarts]{Anwar2012}
Anwar,~M.~S.; Veldhorst,~M.; Brinkman,~A.; Aarts,~J. {Long range supercurrents
  in ferromagnetic CrO$_2$ using a multilayer contact structure}. \emph{Appl.
  Phys. Lett.} \textbf{2012}, \emph{100}, 052602\relax
\mciteBstWouldAddEndPuncttrue
\mciteSetBstMidEndSepPunct{\mcitedefaultmidpunct}
{\mcitedefaultendpunct}{\mcitedefaultseppunct}\relax
\EndOfBibitem
\bibitem[Komori \latin{et~al.}(2021)Komori, Devine-Stoneman, Ohnishi, Yang,
  Devizorova, Mironov, Montiel, {Olde Olthof}, Cohen, Kurebayashi, Blamire,
  Buzdin, and Robinson]{Komori2021}
Komori,~S.; Devine-Stoneman,~J.~M.; Ohnishi,~K.; Yang,~G.; Devizorova,~Z.;
  Mironov,~S.; Montiel,~X.; {Olde Olthof},~L. A.~B.; Cohen,~L.~F.;
  Kurebayashi,~H.; Blamire,~M.~G.; Buzdin,~A.~I.; Robinson,~J. W.~A.
  {Spin-orbit coupling suppression and singlet-state blocking of spin-triplet
  Cooper pairs}. \emph{Sci. Adv.} \textbf{2021}, \emph{7}, eabe0128\relax
\mciteBstWouldAddEndPuncttrue
\mciteSetBstMidEndSepPunct{\mcitedefaultmidpunct}
{\mcitedefaultendpunct}{\mcitedefaultseppunct}\relax
\EndOfBibitem
\bibitem[Iovan \latin{et~al.}(2014)Iovan, Golod, and Krasnov]{Iovan2014}
Iovan,~A.; Golod,~T.; Krasnov,~V.~M. Controllable generation of a spin-triplet
  supercurrent in a Josephson spin valve. \emph{Phys. Rev. B} \textbf{2014},
  \emph{90}, 134514\relax
\mciteBstWouldAddEndPuncttrue
\mciteSetBstMidEndSepPunct{\mcitedefaultmidpunct}
{\mcitedefaultendpunct}{\mcitedefaultseppunct}\relax
\EndOfBibitem
\bibitem[Fominov \latin{et~al.}(2007)Fominov, Volkov, and Efetov]{Fominov2007}
Fominov,~Y.~V.; Volkov,~A.~F.; Efetov,~K.~B. {Josephson effect due to the
  long-range odd-frequency triplet superconductivity in SFS junctions with
  N\'eel domain walls}. \emph{Phys. Rev. B} \textbf{2007}, \emph{75},
  104509\relax
\mciteBstWouldAddEndPuncttrue
\mciteSetBstMidEndSepPunct{\mcitedefaultmidpunct}
{\mcitedefaultendpunct}{\mcitedefaultseppunct}\relax
\EndOfBibitem
\bibitem[Volkov and Efetov(2008)Volkov, and Efetov]{Volkov2008}
Volkov,~A.~F.; Efetov,~K.~B. {Odd triplet superconductivity in a
  superconductor/ferromagnet structure with a narrow domain wall}. \emph{Phys.
  Rev. B} \textbf{2008}, \emph{78}, 024519\relax
\mciteBstWouldAddEndPuncttrue
\mciteSetBstMidEndSepPunct{\mcitedefaultmidpunct}
{\mcitedefaultendpunct}{\mcitedefaultseppunct}\relax
\EndOfBibitem
\bibitem[Kalcheim \latin{et~al.}(2011)Kalcheim, Kirzhner, Koren, and
  Millo]{Kalcheim2011}
Kalcheim,~Y.; Kirzhner,~T.; Koren,~G.; Millo,~O. {Long-range proximity effect
  in La$_{2/3}$Ca$_{1/3}$MnO$_{3}$/(100)YBa$_2$Cu$_3$O$_{7-\delta}$
  ferromagnet/superconductor bilayers: Evidence for induced triplet
  superconductivity in the ferromagnet}. \emph{Phys. Rev. B} \textbf{2011},
  \emph{83}, 064510\relax
\mciteBstWouldAddEndPuncttrue
\mciteSetBstMidEndSepPunct{\mcitedefaultmidpunct}
{\mcitedefaultendpunct}{\mcitedefaultseppunct}\relax
\EndOfBibitem
\bibitem[Aikebaier \latin{et~al.}(2019)Aikebaier, Virtanen, and
  Heikkil{\"{a}}]{Aikebaier2019}
Aikebaier,~F.; Virtanen,~P.; Heikkil{\"{a}},~T. {Superconductivity near a
  magnetic domain wall}. \emph{Phys. Rev. B} \textbf{2019}, \emph{99},
  104504\relax
\mciteBstWouldAddEndPuncttrue
\mciteSetBstMidEndSepPunct{\mcitedefaultmidpunct}
{\mcitedefaultendpunct}{\mcitedefaultseppunct}\relax
\EndOfBibitem
\bibitem[Silaev(2009)]{Silaev2009}
Silaev,~M.~A. Possibility of a long-range proximity effect in a ferromagnetic
  nanoparticle. \emph{Phys. Rev. B} \textbf{2009}, \emph{79}, 184505\relax
\mciteBstWouldAddEndPuncttrue
\mciteSetBstMidEndSepPunct{\mcitedefaultmidpunct}
{\mcitedefaultendpunct}{\mcitedefaultseppunct}\relax
\EndOfBibitem
\bibitem[Kalenkov \latin{et~al.}(2011)Kalenkov, Zaikin, and
  Petrashov]{Kalenkov2011}
Kalenkov,~M.~S.; Zaikin,~A.~D.; Petrashov,~V.~T. Triplet Superconductivity in a
  Ferromagnetic Vortex. \emph{Phys. Rev. Lett.} \textbf{2011}, \emph{107},
  087003\relax
\mciteBstWouldAddEndPuncttrue
\mciteSetBstMidEndSepPunct{\mcitedefaultmidpunct}
{\mcitedefaultendpunct}{\mcitedefaultseppunct}\relax
\EndOfBibitem
\bibitem[Bhatia \latin{et~al.}(2021)Bhatia, Srivastava, Devine-Stoneman,
  Stelmashenko, Barber, Robinson, and Senapati]{Bhatia2021}
Bhatia,~E.; Srivastava,~A.; Devine-Stoneman,~J.; Stelmashenko,~N.~A.;
  Barber,~Z.~H.; Robinson,~J. W.~A.; Senapati,~K. {Nanoscale Domain Wall
  Engineered Spin-Triplet Josephson Junctions and SQUID}. \emph{Nano Lett.}
  \textbf{2021}, \emph{21}, 3092\relax
\mciteBstWouldAddEndPuncttrue
\mciteSetBstMidEndSepPunct{\mcitedefaultmidpunct}
{\mcitedefaultendpunct}{\mcitedefaultseppunct}\relax
\EndOfBibitem
\bibitem[Niu(2012)]{Niu2012}
Niu,~Z. A spin triplet supercurrent in half metal ferromagnet/superconductor
junctions with the interfacial Rashba spin-orbit coupling. \emph{Appl. Phys. Lett.} \textbf{2012}, \emph{101}, 062601\relax
\mciteBstWouldAddEndPuncttrue
\mciteSetBstMidEndSepPunct{\mcitedefaultmidpunct}
{\mcitedefaultendpunct}{\mcitedefaultseppunct}\relax
\EndOfBibitem
\bibitem[Bergeret and Tokatly(2013)Bergeret, and Tokatly]{Bergeret2013_new}
Bergeret,~F.~S.; Tokatly,~I.~V. Singlet-Triplet Conversion and the Long-Range
  Proximity Effect in Superconductor-Ferromagnet Structures with Generic Spin
  Dependent Fields. \emph{Phys. Rev. Lett.} \textbf{2013}, \emph{110},
  117003\relax
\mciteBstWouldAddEndPuncttrue
\mciteSetBstMidEndSepPunct{\mcitedefaultmidpunct}
{\mcitedefaultendpunct}{\mcitedefaultseppunct}\relax
\EndOfBibitem
\bibitem[Bergeret and Tokatly(2014)Bergeret, and Tokatly]{Bergeret2014a}
Bergeret,~F.~S.; Tokatly,~I.~V. Spin-orbit coupling as a source of long-range
  triplet proximity effect in superconductor-ferromagnet hybrid structures.
  \emph{Phys. Rev. B} \textbf{2014}, \emph{89}, 134517\relax
\mciteBstWouldAddEndPuncttrue
\mciteSetBstMidEndSepPunct{\mcitedefaultmidpunct}
{\mcitedefaultendpunct}{\mcitedefaultseppunct}\relax
\EndOfBibitem
\bibitem[Alidoust and Halterman(2015)Alidoust, and
  Halterman]{Alidoust2015_discus1}
Alidoust,~M.; Halterman,~K. {Long-range spin-triplet correlations and edge spin currents in diffusive spin–orbit coupled SNS hybrids with a single spin-active interface}. \emph{New J. Phys} \textbf{2015}, \emph{17}, 033001\relax
\mciteBstWouldAddEndPuncttrue
\mciteSetBstMidEndSepPunct{\mcitedefaultmidpunct}
{\mcitedefaultendpunct}{\mcitedefaultseppunct}\relax
\EndOfBibitem
\bibitem[Jacobsen \latin{et~al.}(2015)Jacobsen, Ouassou, and
  Linder]{Jacobsen2015}
Jacobsen,~S.~H.; Ouassou,~J.~A.; Linder,~J. Critical temperature and tunneling
  spectroscopy of superconductor-ferromagnet hybrids with intrinsic
  Rashba-Dresselhaus spin-orbit coupling. \emph{Phys. Rev. B} \textbf{2015},
  \emph{92}, 024510\relax
\mciteBstWouldAddEndPuncttrue
\mciteSetBstMidEndSepPunct{\mcitedefaultmidpunct}
{\mcitedefaultendpunct}{\mcitedefaultseppunct}\relax
\EndOfBibitem
\bibitem[Satchell and Birge(2018)Satchell, and Birge]{Satchell2018}
Satchell,~N.; Birge,~N.~O. {Supercurrent in ferromagnetic Josephson junctions
  with heavy metal interlayers}. \emph{Phys. Rev. B} \textbf{2018}, \emph{97},
  214509\relax
\mciteBstWouldAddEndPuncttrue
\mciteSetBstMidEndSepPunct{\mcitedefaultmidpunct}
{\mcitedefaultendpunct}{\mcitedefaultseppunct}\relax
\EndOfBibitem
\bibitem[Jeon \latin{et~al.}(2018)Jeon, Ciccarelli, Ferguson, Kurebayashi,
  Cohen, Montiel, Eschrig, Robinson, and Blamire]{Jeon2018}
Jeon,~K.~R.; Ciccarelli,~C.; Ferguson,~A.~J.; Kurebayashi,~H.; Cohen,~L.~F.;
  Montiel,~X.; Eschrig,~M.; Robinson,~J.~W.; Blamire,~M.~G. {Enhanced spin
  pumping into superconductors provides evidence for superconducting pure spin
  currents}. \emph{Nat. Mater.} \textbf{2018}, \emph{17}, 499\relax
\mciteBstWouldAddEndPuncttrue
\mciteSetBstMidEndSepPunct{\mcitedefaultmidpunct}
{\mcitedefaultendpunct}{\mcitedefaultseppunct}\relax
\EndOfBibitem
\bibitem[Bujnowski \latin{et~al.}(2019)Bujnowski, Biele, and
  Bergeret]{Bujnowski2019}
Bujnowski,~B.; Biele,~R.; Bergeret,~F.~S. {Switchable Josephson current in
  junctions with spin-orbit coupling}. \emph{Phys. Rev. B} \textbf{2019},
  \emph{100}, 224518\relax
\mciteBstWouldAddEndPuncttrue
\mciteSetBstMidEndSepPunct{\mcitedefaultmidpunct}
{\mcitedefaultendpunct}{\mcitedefaultseppunct}\relax
\EndOfBibitem
\bibitem[Eskilt \latin{et~al.}(2019)Eskilt, Amundsen, Banerjee, and
  Linder]{Eskilt2019}
Eskilt,~J.~R.; Amundsen,~M.; Banerjee,~N.; Linder,~J. Long-ranged triplet
  supercurrent in a single in-plane ferromagnet with spin-orbit coupled
  contacts to superconductors. \emph{Phys. Rev. B} \textbf{2019}, \emph{100},
  224519\relax
\mciteBstWouldAddEndPuncttrue
\mciteSetBstMidEndSepPunct{\mcitedefaultmidpunct}
{\mcitedefaultendpunct}{\mcitedefaultseppunct}\relax
\EndOfBibitem
\bibitem[Jeon \latin{et~al.}(2020)Jeon, Montiel, Komori, Ciccarelli, Haigh,
  Kurebayashi, Cohen, Chan, Stenning, Lee, Blamire, and Robinson]{Jeon2020}
Jeon,~K.~R.; Montiel,~X.; Komori,~S.; Ciccarelli,~C.; Haigh,~J.;
  Kurebayashi,~H.; Cohen,~L.~F.; Chan,~A.~K.; Stenning,~K.~D.; Lee,~C.~M.;
  Blamire,~M.~G.; Robinson,~J.~W. {Tunable Pure Spin Supercurrents and the
  Demonstration of Their Gateability in a Spin-Wave Device}. \emph{Phys. Rev.
  X} \textbf{2020}, \emph{10}, 31020\relax
\mciteBstWouldAddEndPuncttrue
\mciteSetBstMidEndSepPunct{\mcitedefaultmidpunct}
{\mcitedefaultendpunct}{\mcitedefaultseppunct}\relax
\EndOfBibitem
\bibitem[Satchell \latin{et~al.}(2019)Satchell, Loloee, and
  Birge]{Satchell2019}
Satchell,~N.; Loloee,~R.; Birge,~N.~O. Supercurrent in ferromagnetic Josephson
  junctions with heavy-metal interlayers. II. Canted magnetization. \emph{Phys.
  Rev. B} \textbf{2019}, \emph{99}, 174519\relax
\mciteBstWouldAddEndPuncttrue
\mciteSetBstMidEndSepPunct{\mcitedefaultmidpunct}
{\mcitedefaultendpunct}{\mcitedefaultseppunct}\relax
\EndOfBibitem
\bibitem[Silaev \latin{et~al.}(2020)Silaev, Bobkova, and Bobkov]{Silaev2020}
Silaev,~M.~A.; Bobkova,~I.~V.; Bobkov,~A.~M. Odd triplet superconductivity
  induced by a moving condensate. \emph{Phys. Rev. B} \textbf{2020},
  \emph{102}, 100507\relax
\mciteBstWouldAddEndPuncttrue
\mciteSetBstMidEndSepPunct{\mcitedefaultmidpunct}
{\mcitedefaultendpunct}{\mcitedefaultseppunct}\relax
\EndOfBibitem
\bibitem[Tokatly \latin{et~al.}(2019)Tokatly, Bujnowski, and
  Bergeret]{Tokatly2019}
Tokatly,~I.~V.; Bujnowski,~B.; Bergeret,~F.~S. {Universal correspondence
  between edge spin accumulation and equilibrium spin currents in nanowires
  with spin-orbit coupling}. \emph{Phys. Rev. B} \textbf{2019}, \emph{100},
  214422\relax
\mciteBstWouldAddEndPuncttrue
\mciteSetBstMidEndSepPunct{\mcitedefaultmidpunct}
{\mcitedefaultendpunct}{\mcitedefaultseppunct}\relax
\EndOfBibitem
\bibitem[Bobkova and Barash(2004)Bobkova, and Barash]{bobkova2004effects}
Bobkova,~I.~V.; Barash,~Y.~S. Effects of spin-orbit interaction on
  superconductor-ferromagnet heterostructures: Spontaneous electric and spin
  surface currents. \emph{J. Exp. Theor.} \textbf{2004}, \emph{80},
  494\relax
\mciteBstWouldAddEndPuncttrue
\mciteSetBstMidEndSepPunct{\mcitedefaultmidpunct}
{\mcitedefaultendpunct}{\mcitedefaultseppunct}\relax
\EndOfBibitem
\bibitem[Alidoust and Halterman(2015)Alidoust, and
  Halterman]{Alidoust2015_discus2}
Alidoust,~M.; Halterman,~K. {Long-range spin-triplet correlations and edge spin
  currents in diffusive spin–orbit coupled SNS hybrids with a single
  spin-active interface}. \emph{J. Condens. Matter Phys.} \textbf{2015},
  \emph{27}, 235301\relax
\mciteBstWouldAddEndPuncttrue
\mciteSetBstMidEndSepPunct{\mcitedefaultmidpunct}
{\mcitedefaultendpunct}{\mcitedefaultseppunct}\relax
\EndOfBibitem
\bibitem[Salamone \latin{et~al.}(2021)Salamone, Svendsen, Amundsen, and
  Jacobsen]{Salamone2021}
Salamone,~T.; Svendsen,~M. B.~M.; Amundsen,~M.; Jacobsen,~S. {Curvature-induced
  long ranged supercurrents in diffusive SFS Josephson Junctions, with dynamic
  $0-\pi$ transition}. \emph{Phys. Rev. B} \textbf{2021}, \emph{104},
  L060505\relax
\mciteBstWouldAddEndPuncttrue
\mciteSetBstMidEndSepPunct{\mcitedefaultmidpunct}
{\mcitedefaultendpunct}{\mcitedefaultseppunct}\relax
\EndOfBibitem
\bibitem[Mazanik and Bobkova(2022)Mazanik, and Bobkova]{Mazanik2022}
Mazanik,~A.~A.; Bobkova,~I.~V. Supercurrent-induced long-range triplet
  correlations and controllable Josephson effect in superconductor/ferromagnet
  hybrids with extrinsic spin-orbit coupling. \emph{Phys. Rev. B}
  \textbf{2022}, \emph{105}, 144502\relax
\mciteBstWouldAddEndPuncttrue
\mciteSetBstMidEndSepPunct{\mcitedefaultmidpunct}
{\mcitedefaultendpunct}{\mcitedefaultseppunct}\relax
\EndOfBibitem
\bibitem[Bobkova \latin{et~al.}(2021)Bobkova, Bobkov, and
  Silaev]{Bobkova2021_new}
Bobkova,~I.~V.; Bobkov,~A.~M.; Silaev,~M.~A. Dynamic Spin-Triplet Order Induced
  by Alternating Electric Fields in Superconductor-Ferromagnet-Superconductor
  Josephson Junctions. \emph{Phys. Rev. Lett.} \textbf{2021}, \emph{127},
  147701\relax
\mciteBstWouldAddEndPuncttrue
\mciteSetBstMidEndSepPunct{\mcitedefaultmidpunct}
{\mcitedefaultendpunct}{\mcitedefaultseppunct}\relax
\EndOfBibitem
\bibitem[Dynes and Fulton(1971)Dynes, and Fulton]{Dynes1971}
Dynes,~R.~C.; Fulton,~T.~A. Supercurrent Density Distribution in Josephson
  Junctions. \emph{Phys. Rev. B} \textbf{1971}, \emph{3}, 3015\relax
\mciteBstWouldAddEndPuncttrue
\mciteSetBstMidEndSepPunct{\mcitedefaultmidpunct}
{\mcitedefaultendpunct}{\mcitedefaultseppunct}\relax
\EndOfBibitem
\bibitem[Suominen \latin{et~al.}(2017)Suominen, Danon, Kjaergaard, Flensberg,
  Shabani, Palmstr{\o}m, Nichele, and Marcus]{Suominen2017}
Suominen,~H.~J.; Danon,~J.; Kjaergaard,~M.; Flensberg,~K.; Shabani,~J.;
  Palmstr{\o}m,~C.~J.; Nichele,~F.; Marcus,~C.~M. {Anomalous Fraunhofer
  interference in epitaxial superconductor-semiconductor Josephson junctions}.
  \emph{Phys. Rev. B} \textbf{2017}, \emph{95}, 035307\relax
\mciteBstWouldAddEndPuncttrue
\mciteSetBstMidEndSepPunct{\mcitedefaultmidpunct}
{\mcitedefaultendpunct}{\mcitedefaultseppunct}\relax
\EndOfBibitem
\bibitem[Huang \latin{et~al.}(2019)Huang, Zhou, Zhang, Yang, Liu, Wang, Wan,
  Huang, Liao, Zhang, Liu, Deng, Chen, Han, Zou, Lin, Han, Wang, Law, and
  Xiu]{Huang2019}
Huang,~C. \latin{et~al.}  {Proximity-induced surface superconductivity in Dirac
  semimetal Cd$_3$As$_2$}. \emph{Nat. Commun.} \textbf{2019}, \emph{10},
  2217\relax
\mciteBstWouldAddEndPuncttrue
\mciteSetBstMidEndSepPunct{\mcitedefaultmidpunct}
{\mcitedefaultendpunct}{\mcitedefaultseppunct}\relax
\EndOfBibitem
\bibitem[Smilde \latin{et~al.}(2002)Smilde, Ariando, Blank, Gerritsma,
  Hilgenkamp, and Rogalla]{zigzag1}
Smilde,~H. J.~H.; Ariando,; Blank,~D. H.~A.; Gerritsma,~G.~J.; Hilgenkamp,~H.;
  Rogalla,~H. {d-Wave--Induced josephson current counterflow in
  $\text{YBa}_{2}\text{Cu}_{3}\text{O}_{7}/\text{Nb}$ zigzag junctions}.
  \emph{Phys. Rev. Lett.} \textbf{2002}, \emph{88}, 057004\relax
\mciteBstWouldAddEndPuncttrue
\mciteSetBstMidEndSepPunct{\mcitedefaultmidpunct}
{\mcitedefaultendpunct}{\mcitedefaultseppunct}\relax
\EndOfBibitem
\bibitem[Scharinger \latin{et~al.}(2010)Scharinger, G\"urlich, Mints, Weides,
  Kohlstedt, Goldobin, Koelle, and Kleiner]{zigzag2}
Scharinger,~S.; G\"urlich,~C.; Mints,~R.~G.; Weides,~M.; Kohlstedt,~H.;
  Goldobin,~E.; Koelle,~D.; Kleiner,~R. Interference patterns of multifacet
  $20\ifmmode\times\else\texttimes\fi{}(0\text{\ensuremath{-}}\ensuremath{\pi})$
  Josephson junctions with ferromagnetic barrier. \emph{Phys. Rev. B}
  \textbf{2010}, \emph{81}, 174535\relax
\mciteBstWouldAddEndPuncttrue
\mciteSetBstMidEndSepPunct{\mcitedefaultmidpunct}
{\mcitedefaultendpunct}{\mcitedefaultseppunct}\relax
\EndOfBibitem
\bibitem[G\"urlich \latin{et~al.}(2010)G\"urlich, Scharinger, Weides,
  Kohlstedt, Mints, Goldobin, Koelle, and Kleiner]{zigzag3}
G\"urlich,~C.; Scharinger,~S.; Weides,~M.; Kohlstedt,~H.; Mints,~R.~G.;
  Goldobin,~E.; Koelle,~D.; Kleiner,~R. Visualizing supercurrents in
  ferromagnetic Josephson junctions with various arrangements of 0 and
  $\ensuremath{\pi}$ segments. \emph{Phys. Rev. B} \textbf{2010}, \emph{81},
  094502\relax
\mciteBstWouldAddEndPuncttrue
\mciteSetBstMidEndSepPunct{\mcitedefaultmidpunct}
{\mcitedefaultendpunct}{\mcitedefaultseppunct}\relax
\EndOfBibitem
\bibitem[Bergeret and Tokatly(2020)Bergeret, and Tokatly]{bergeret2020}
Bergeret,~F.~S.; Tokatly,~I.~V. Theory of the magnetic response in finite
  two-dimensional superconductors. \emph{Phys. Rev. B} \textbf{2020},
  \emph{102}, 060506(R)\relax
\mciteBstWouldAddEndPuncttrue
\mciteSetBstMidEndSepPunct{\mcitedefaultmidpunct}
{\mcitedefaultendpunct}{\mcitedefaultseppunct}\relax
\EndOfBibitem
\bibitem[Tokatly(2008)]{tokatly2008equilibrium}
Tokatly,~I. Equilibrium spin currents: non-abelian gauge invariance and color
  diamagnetism in condensed matter. \emph{Phys. Rev. Lett.} \textbf{2008},
  \emph{101}, 106601\relax
\mciteBstWouldAddEndPuncttrue
\mciteSetBstMidEndSepPunct{\mcitedefaultmidpunct}
{\mcitedefaultendpunct}{\mcitedefaultseppunct}\relax
\EndOfBibitem
\bibitem[Hill \latin{et~al.}(2021)Hill, Slastikov, and
  Tchernyshyov]{10.21468/SciPostPhys.10.3.078}
Hill,~D.; Slastikov,~V.; Tchernyshyov,~O. {Chiral magnetism: a geometric
  perspective}. \emph{SciPost Phys.} \textbf{2021}, \emph{10}, 78\relax
\mciteBstWouldAddEndPuncttrue
\mciteSetBstMidEndSepPunct{\mcitedefaultmidpunct}
{\mcitedefaultendpunct}{\mcitedefaultseppunct}\relax
\EndOfBibitem
\bibitem[Bass and Pratt(2007)Bass, and Pratt]{Bass_2007}
Bass,~J.; Pratt,~W.~P. Spin-diffusion lengths in metals and alloys, and
  spin-flipping at metal/metal interfaces: an experimentalist's critical
  review. \emph{J. Condens. Matter Phys.} \textbf{2007}, \emph{19},
  183201\relax
\mciteBstWouldAddEndPuncttrue
\mciteSetBstMidEndSepPunct{\mcitedefaultmidpunct}
{\mcitedefaultendpunct}{\mcitedefaultseppunct}\relax
\EndOfBibitem
\bibitem[Glick \latin{et~al.}(2018)Glick, Aguilar, Gougam, Niedzielski,
  Gingrich, Loloee, Pratt, and Birge]{Glick2018}
Glick,~J.~A.; Aguilar,~V.; Gougam,~A.~B.; Niedzielski,~B.~M.; Gingrich,~E.~C.;
  Loloee,~R.; Pratt,~W.~P.; Birge,~N.~O. {Phase control in a spin-triplet
  SQUID}. \emph{Sci. Adv.} \textbf{2018}, \emph{4}, eaat9457\relax
\mciteBstWouldAddEndPuncttrue
\mciteSetBstMidEndSepPunct{\mcitedefaultmidpunct}
{\mcitedefaultendpunct}{\mcitedefaultseppunct}\relax
\EndOfBibitem
\end{mcitethebibliography}


\providecommand{\latin}[1]{#1}
\makeatletter
\providecommand{\doi}
  {\begingroup\let\do\@makeother\dospecials
  \catcode`\{=1 \catcode`\}=2 \doi@aux}
\providecommand{\doi@aux}[1]{\endgroup\texttt{#1}}
\makeatother
\providecommand*\mcitethebibliography{\thebibliography}
\csname @ifundefined\endcsname{endmcitethebibliography}
  {\let\endmcitethebibliography\endthebibliography}{}


\end{document}


\title{Supporting Information for: Superconducting triplet rim currents in a spin-textured ferromagnetic disk}

\date{\today}

\author{Remko Fermin}
\author{Dyon van Dinter}
\author{Michel Hubert}
\author{Bart Woltjes}
\affiliation{Huygens-Kamerlingh Onnes Laboratory, Leiden University, P.O. Box 9504, 2300 RA Leiden, The Netherlands.}
\author{Mikhail Silaev}
\affiliation{Department of Physics and Nanoscience Center, University of Jyv\"askyl\"a, P.O.
Box 35 (YFL), FI-40014 University of Jyv\"askyl\"a, Finland}
\affiliation{Computational Physics Laboratory, Physics Unit, Faculty of Engineering and Natural Sciences, Tampere University, P.O. Box 692, Tampere, Finland}
\author{Jan Aarts}
\author{Kaveh Lahabi}
\email{lahabi@physics.leidenuniv.nl}
\affiliation{Huygens-Kamerlingh Onnes Laboratory, Leiden University, P.O. Box 9504, 2300 RA Leiden, The Netherlands.}

\pacs{} \maketitle

\section{Device fabrication}

A four-probe contact geometry is patterned on Si substrates, using electron-beam lithography. A cobalt (60 nm) niobium (45 nm) bilayer was deposited in an ultra-high vacuum by Ar-sputtering. The thickness of the Co layer ensures a stable vortex magnetization for the disk diameters discussed in this work (1.05 $\mu$m and 1.62 $\mu$m). The Co weak link is formed by a line cut through the Nb layer, using an ultra-low beam current of 1.5 pA. The resulting trench separates the Nb electrodes by a $\sim 20$ nm Co weak link at the center of the disk. A false-colored electron micrograph of a $1.05$ $\mu$m diameter device is shown in Figure 1b of the main text. To establish a possible link between spin-diffusion length and the width of the rim current channels, several samples, including the one used for Figure 4 of the main text, were deposited with a copper (5 nm) buffer layer, inserted in between the Nb and Co (i.e., after FIB structuring, the layer configuration of the planar junction was Nb/Cu/Co/Cu/Nb). The Cu layer was introduced to enhance the spin-diffusion length at the S-F interface. However, the limited number of lobes in the interference patterns of such devices prevents the Fourier analysis from providing the necessary spatial resolution to quantify the variations in the size of the rim current channel. Qualitatively, no discernible changes could be attributed to the presence of the Cu buffer layer below the Nb electrodes.

\section{Micromagnetic simulations}

A full description of the micromagnetic simulations of the cobalt layer is reported in a previous work.\cite{Lahabi2017a}  Within the Object Oriented MicroMagnetic Framework (OOMMF) simulation software, we model the multilayer by dividing it into a three-dimensional mesh of 5 nm cubic cells. The exchange coefficient and saturation magnetization of Co are 3 $\times~10^{-11}$ J/m and 1.4  $\times~10^{-6}$ A/m respectively. The Gilbert damping constant $\alpha$ was set to 0.5 to allow for rapid convergence. In order to represent the polycrystalline nature of the sputtered films, we define the direction of anisotropy by a random vector field.

\section{Control experiments}\label{control}

In this section we establish the vital role of spin texture for generating long range triplet (LRT) correlations. Furthermore, we show that junctions, where transport is dominated by singlet correlations, are not affected by the spin texture and that the two-channel behavior is completely absent.

\begin{figure*}[t!]
 \centerline{$
 \begin{array}{c}
  \includegraphics[width=1\linewidth]{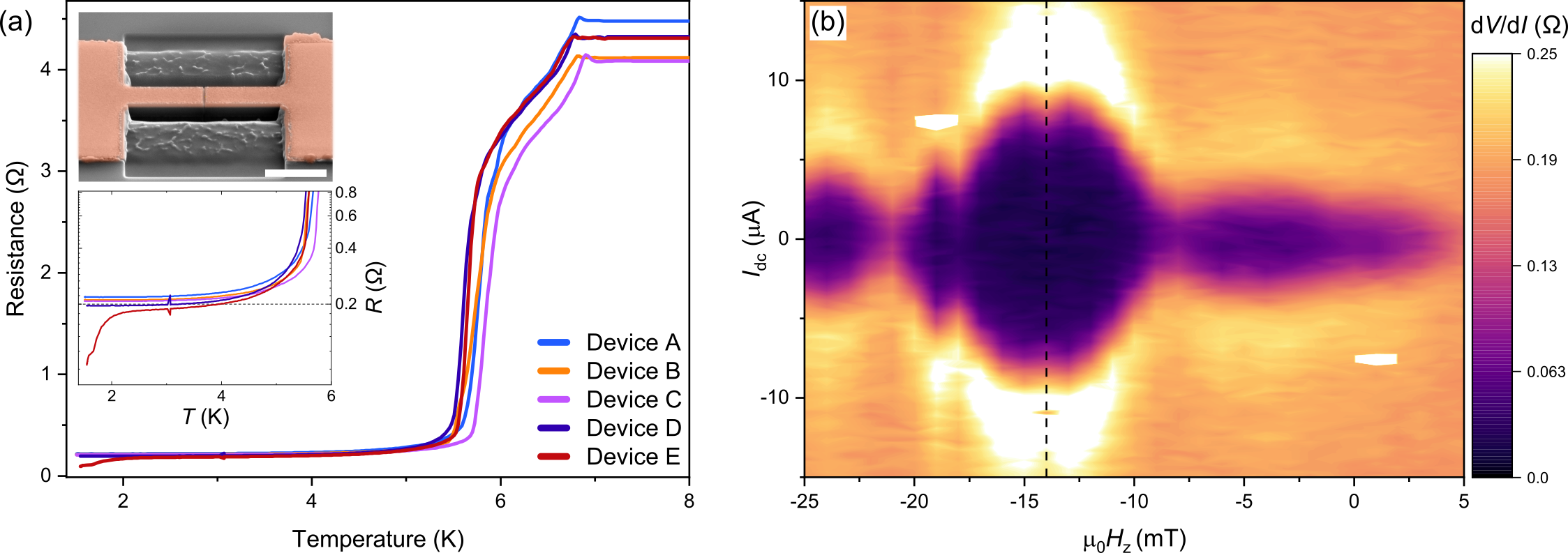}
 \end{array}$}
 \caption{\textbf{a.} Resistance vs temperature obtained on five bar junction devices lacking any spin texture, using a measurement current of 10 $\mu$A. The weak link in devices A to D is fabricated following the same method as the triplet disk devices. The low T resistance of these devices equals the weak link resistance (0.2 $\Omega$). As can be seen from the lower inset, only device E shows weak signs of proximity effect, with a heavily suppressed critical current. This device was fabricated using a lower Ga-ion dose, resulting in a Nb-residue in the weak link. The top inset shows a false-colored electron micrograph of a bar junction where the scale bar corresponds to 2 $\mu$m. \textbf{b.} shows the SQI pattern obtained on device E. The Fraunhofer pattern indicates a uniform current distribution throughout the weak link. The center lobe is shifted to -14 mT due to the magnetic stray fields of the ferromagnet.} \label{FIGS1}
\end{figure*}

We fabricated control samples with rectangular configuration, where the Co layer is uniformly magnetized along its long axis (\textit{i.e.}, has no spin texture). The bar-shaped devices have equal width (1.05 $\mu$m) and layer thicknesses as the triplet disk devices (see top inset of Figure \ref{FIGS1}a). Due to the 5:1 aspect ratio, no magnetic vortex is stable in the cobalt layer. The same focused ion beam milling procedure was applied to form the weak link. Figure \ref{FIGS1}a shows the resistance vs temperature of five of such junctions. For devices A to D a similar Ga-ion dose is used for preparing the trench as the triplet disk devices (Nb is completely removed from the weak link). In device E, however, the weak link is fabricated by applying a lower dose, which results in the incomplete removal of the Nb layer in the trench. As shown in Figure \ref{FIGS1}a, only device E shows signs of short-range proximity effect, with a heavily suppressed critical current. Figure \ref{FIGS1}b shows the superconducting quantum interference (SQI) pattern of device E. In contrast to the triplet disk junctions, we observe a distorted Fraunhofer pattern with a rapidly decaying $I_\text{c}(B)$, \textit{i.e.}, no two-channel behavior. The pattern is shifted and distorted due to the magnetic dipole fields of the ferromagnet.

\begin{figure*}[bt!]
 \centerline{$
 \begin{array}{c}
  \includegraphics[width=0.85\linewidth]{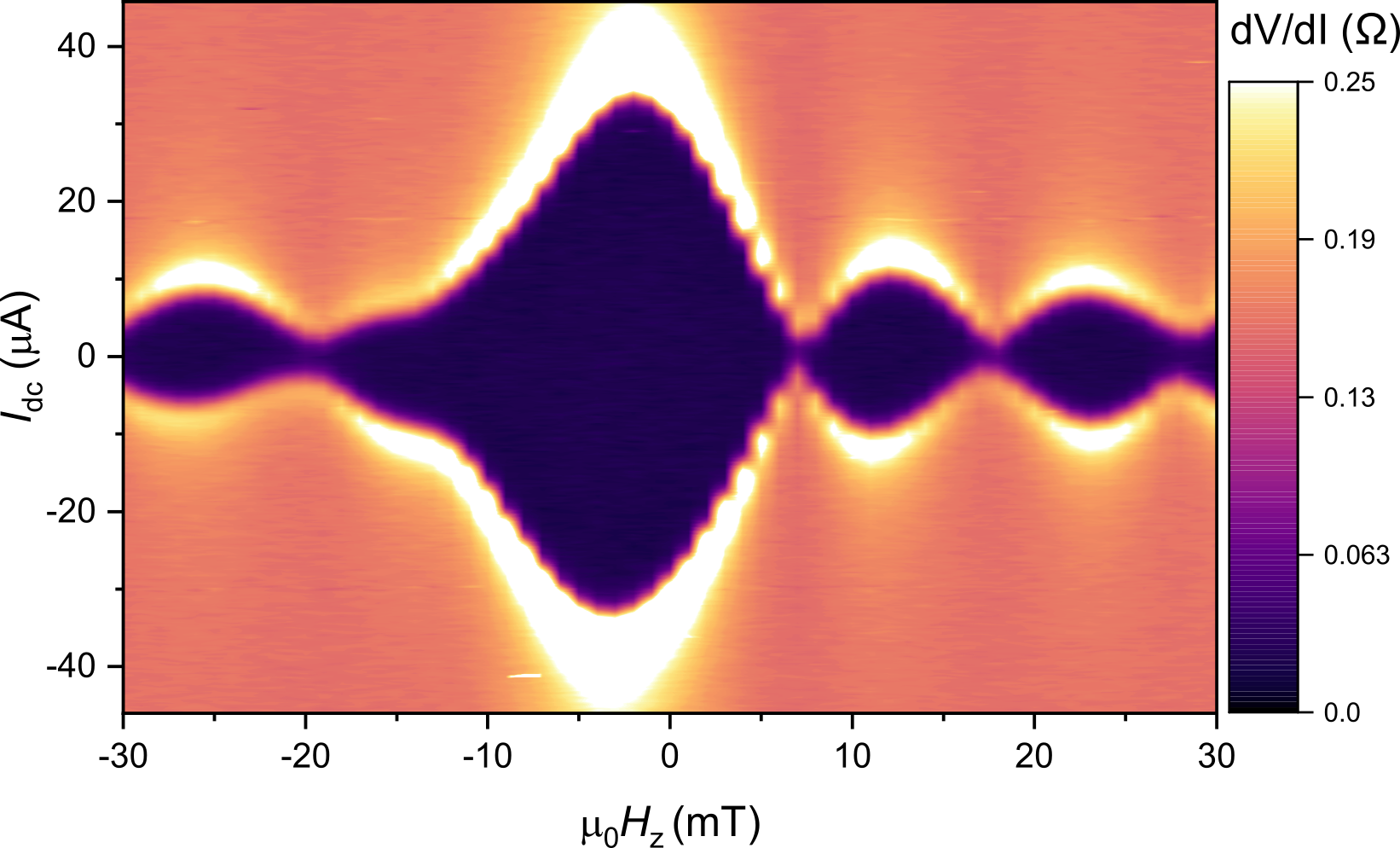}
 \end{array}$}
 \caption{SQI pattern recorded on a disk-shaped sample containing a shallow trench (\textit{i.e.}, with a Nb-residue in the weak link) and therefore dominated by singlet transport. The central peak is twice as wide as neighboring ones, and the peak height decreases in a 1/$B$ fashion. Superconductivity in this sample is found to be unaffected by the removal of the spin texture of the ferromagnet by the application of an in-plane field.} \label{FIGS0}
\end{figure*}

\begin{figure*}[bt!]
 \centerline{$
 \begin{array}{c}
  \includegraphics[width=1\linewidth]{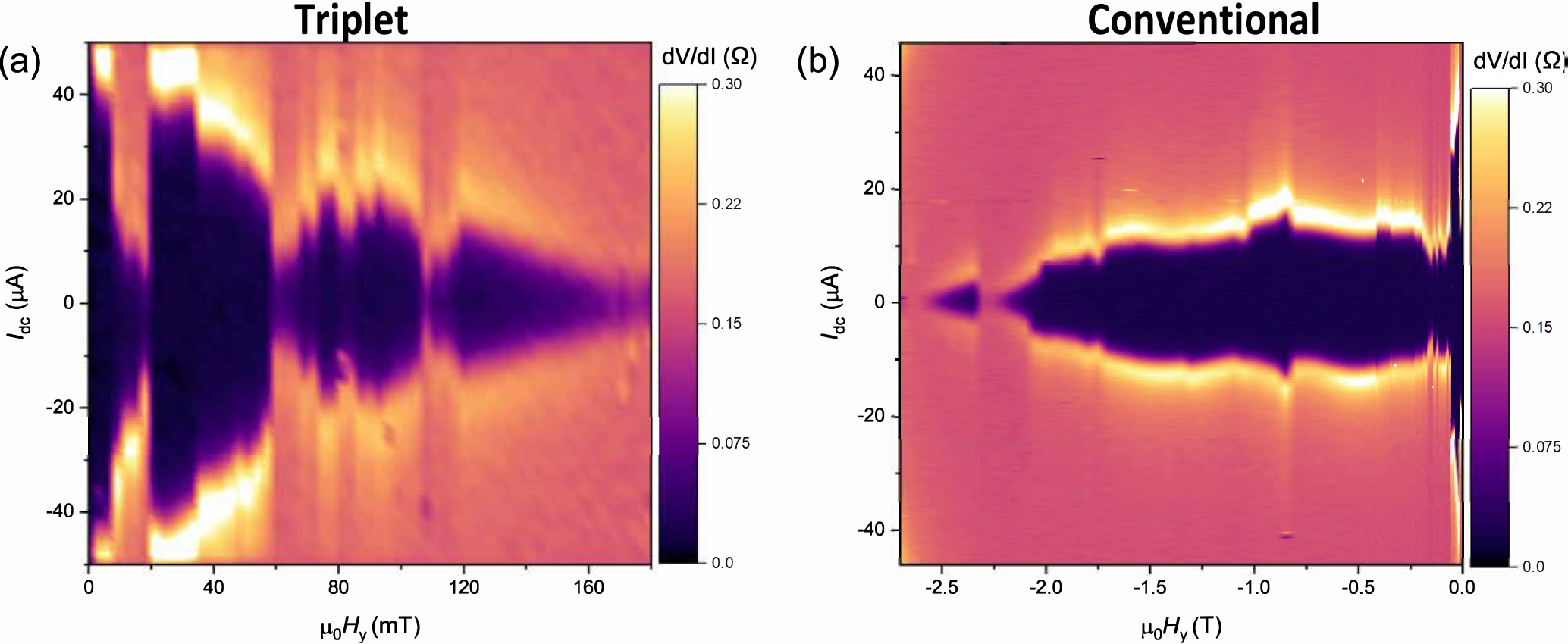}
 \end{array}$}
 \caption{In-plane $I_\text{c}(H_y)$ for a sample with triplet channels and a sample with a conventional weak link. \textbf{a.} The triplet sample loses its critical current above 180~mT in-plane field along the trench. At these fields, the spin texture of the disk has become uniform and therefore the generator for LRT correlations is absent. On the sample in \textbf{a.} we obtained the data shown in Figure 4 of the main text. \textbf{b.} The sample with a shallow trench retains a $I_\text{c}$ up to above 2 T, far above the fields required for stabilizing a uniform spin-texture.}\label{REB2}
\end{figure*}

We additionally fabricated disk-shaped control samples where the Nb is not completely removed from the weak link to study the interaction of singlet supercurrents with the vortex spin texture. Figure \ref{FIGS0} shows an SQI pattern of such a device. Contrary to the triplet devices, we observe a clear Fraunhofer SQI pattern with the central lobe being twice as wide as the subsequent ones and a 1/$B$ decay of the peak height. The missing minimum on the negative field side of the pattern may be attributed to the presence of a nearby Abrikosov vortex in one of the superconducting leads.\cite{Golod2019} This demonstrates that, despite the ferromagnetic vortex in the Co-layer, the singlet supercurrent is distributed uniformly across the junction. It should also be noted that conventional weak link devices maintain their critical current, even when the spin texture is removed by applying an in-plane field. In Figure~ \ref{REB2} we compare the behavior of a device with triplet channels to that of a device with a shallow trench, i.e., a conventional weak link, under influence of an in-plane field $H_y$ along the trench. The triplet device shows variations in $I_c$ due to 0-$\pi$ transitions and vice versa, and disappearance of the critical current around 180~mT. The conventional device containing the shallow trench (Figure \ref{FIGS0}) retains an $I_c$ of around 10~$\mu$A up to 2~T.

\begin{figure*}[bt!]
 \centerline{$
 \begin{array}{c}
  \includegraphics[width=0.95\linewidth]{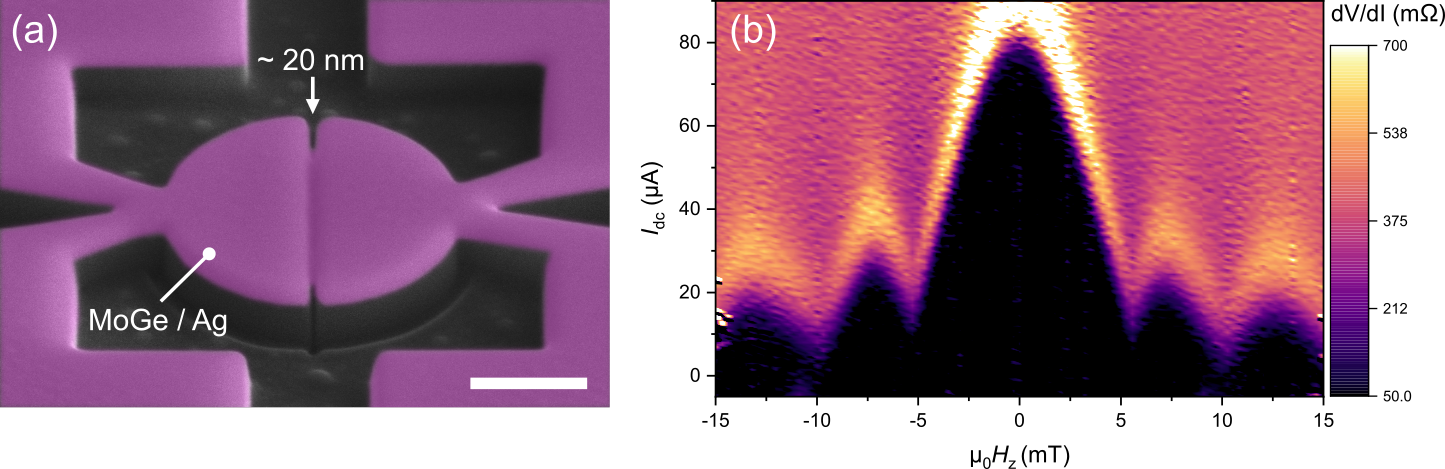}
 \end{array}$}
 \caption{disk-shaped sample milled out of a Ag/MoGe bilayer and its corresponding SQI pattern. \textbf{a.} A false colored electron micrograph of a disk-shaped sample fabricated to have a normal metal weak link. This sample is therefore characterized by singlet transport only. The scale bar corresponds to 500 nm. The SQI pattern displayed in \textbf{b.} is a clear Fraunhofer pattern indicating a uniform current distribution, like the case of the shallow trench devices. The pattern is found to be unaffected by the application of in-plane fields.} \label{FIGS_normal}
\end{figure*}

Finally, in order to verify that the two-channel behavior is not a trivial consequence of the disk configuration, or a byproduct of the FIB milling used in forming the weak link, we  examine the Josephson transport of disk-shaped junctions with a non-magnetic barrier. These control samples are structured from a MoGe (55 nm)/Ag (20 nm) bilayer. The Ag weak link is formed through the same FIB treatment as the one applied to the Nb/Co/Nb junctions discussed in the main text. A false colored electron micrograph is shown in Figure \ref{FIGS_normal}a and Figure \ref{FIGS_normal}b displays an SQI pattern of such device; we find a typical and undistorted Fraunhofer pattern, corresponding to a uniform distribution of supercurrent in the junction. Contrarily to the SF-devices presented in the main text, the shape of the pattern remains unaffected by the application of in-plane fields. To summarize, our control experiments demonstrate that the rim channels only appear when transport is carried by the LRT correlations. In contrast to singlet or SRT transport, the LRT currents are also highly sensitive to the changes in cobalt spin texture, which are brought about by relatively small in-plane fields.

\section{Extracting the supercurrent density with Fourier analysis}

In their 1971 paper, Dynes and Fulton~\cite{Dynes1971} discuss a relation between the shape of the SQI pattern and the supercurrent distribution in a junction. They realized that, in the case of sinusoidal current phase relation,\footnote{The disk devices have a sinusoidal current phase relation since they are in the short junction limit.} the current distribution in the junction can be extracted by complex inverse Fourier transform of the field dependence of the critical current. Since the supercurrent can be found from integrating the critical current density along the width of the junction, we can write:

\begin{align} \label{EqS1}
I_s(\beta,\phi) =  \int_{-W/2}^{W/2} J_s(w) \text{d}w = \int_{-W/2}^{W/2} J_C(w) \sin{(\beta w+\phi)} \: \text{d}w = \operatorname{Im} \left( e^{i\phi}\int_{-\infty}^{\infty} J_C(w) e^{i\beta w} \: \text{d}w \right)
\end{align}

\noindent Here $w$ is the position along the junction, $W$ the junction width, $\beta = 2\pi L_{\text{eff}} B / \Phi_0$ is the normalized field ($L_{\text{eff}}$ the effective junction length, $B$ the applied field and $\Phi_0 = h/2e$) and $\phi$ the phase difference at the center of the junction due to a voltage bias to the sample. In the last part, the integration bounds have been extended to infinity since outside the sample there are no supercurrents. The critical current is given by the absolute value of the complex expression:

\begin{align} \label{EqS2}
I_c(\beta) =   \left| \int_{-\infty}^{\infty} J_C(w) e^{i\beta w} \: \text{d}w \right| = \left| \Im_c \right| 
\end{align}

\noindent From this expression a Fourier transform can be recognized where position along the junctions width $w$ and $\beta$ form conjugate variables. The transform $\Im_c$ is complex and therefore its real and imaginary parts encode for the even and odd components in $I_c(\beta)$ respectively. Since the interference patterns measured in our experiments are relatively symmetric (\textit{i.e.}, an even function of the applied magnetic field), we can assume $\Im_c$ to be dominantly real:

\begin{align} \label{EqS3}
I_{c,\text{even}}(\beta) =   \int_{-\infty}^{\infty} J_{c,\text{even}}(w) \cos(\beta w)  \: \text{d}w
\end{align}

\noindent Therefore the real part of $\Im_c$ is an oscillating function that flips sign at each zero crossing. The imaginary part is expected to be significantly smaller than the real part, except at the zero-crossing where the even part vanishes. Therefore, the imaginary part of $\Im_c$ ($I_{c,\text{odd}}(\beta)$) can be approximated by the critical current at the minima, flipping sign between each minimum and linearly interpolating between them. The inverse transform yielding $J_C(w)$ from $I_c(\beta)$ is then given by:

\begin{align} \label{EqS4}
J_C(w) =  \left|  \int_{-\infty}^{\infty} \left( I_{c,\text{even}}(\beta) + i I_{c,\text{odd}}(\beta) \right) e^{i\beta w} \: \text{d}\beta \right| = \left|  \int_{-\infty}^{\infty} \Im_c \: e^{i\beta w} \: \text{d}\beta \right|
\end{align}

\begin{figure*}[bt!]
 \centerline{$
 \begin{array}{c}
  \includegraphics[width=1\linewidth]{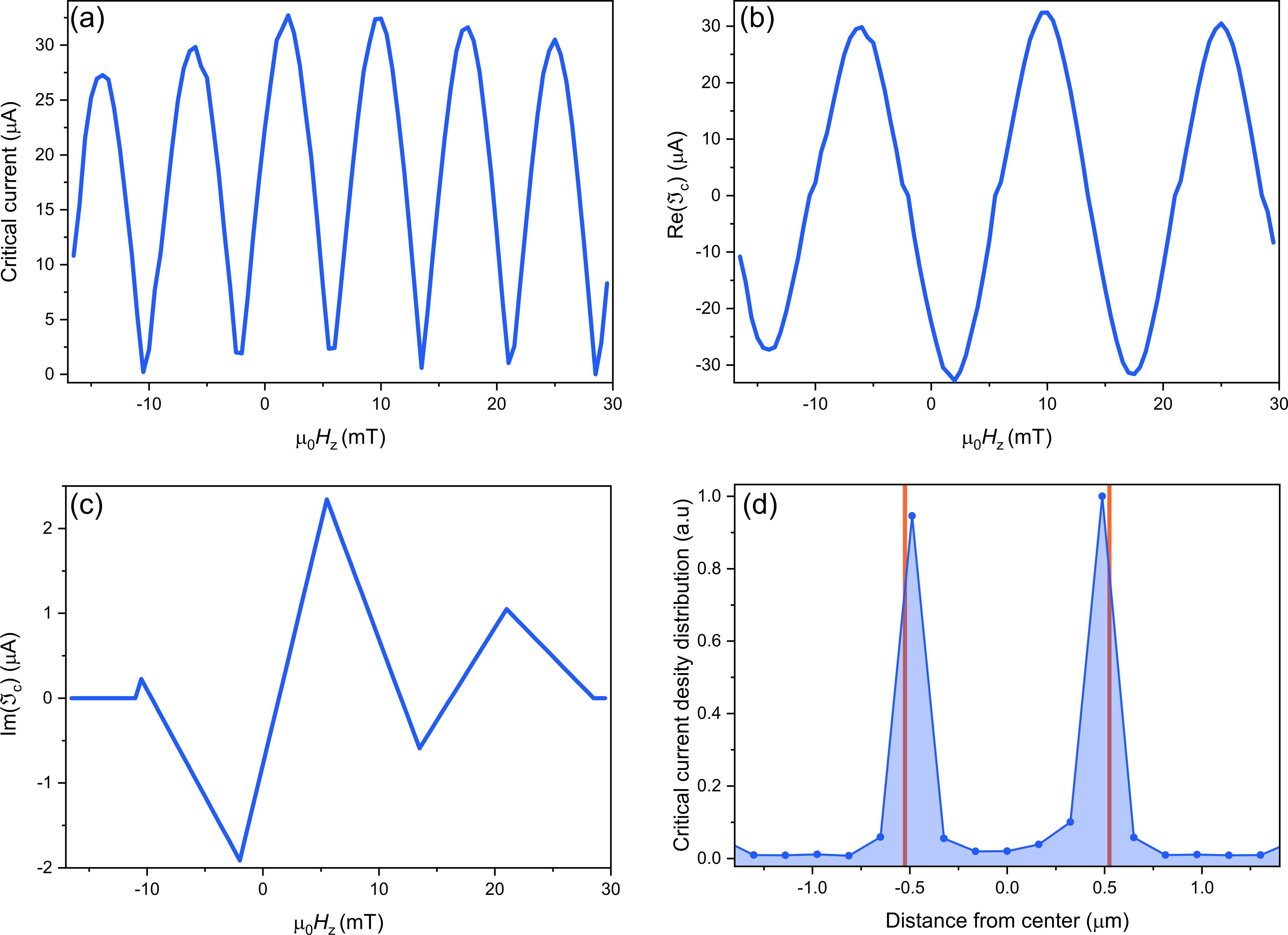}
 \end{array}$}
 \caption{\textbf{a.} Extracted critical current on basis of a voltage cut-off. The global minimum of the interference pattern is shifted to zero current, to not overestimate the imaginary part of the complex critical current. \textbf{b.} and \textbf{c.} respectively the real and imaginary part of the complex critical current. These are extracted from the interference pattern depicted in a. following the procedure from the main text. \textbf{d.} The resulting critical current density distribution calculated using the Fourier transform described in Eq. \ref{EqS4} from the data in b. and c.} \label{FIGS2}
\end{figure*}

\noindent We extract the critical current by a voltage cut-off in each subsequent $IV$-measurement (\textit{i.e.}, $I_c$ is the current for which we measure a certain voltage over the junction). We observe the minima of the SQI pattern to be non-zero, which might be originating from asymmetries in loop induction or junction capacitance between the two current channels. We vertically translate the extracted $I_c$ values such that the global minimum equals zero current (see figure \ref{FIGS2}a). This step in the data analysis prevents the overestimation of $I_{c,\text{odd}}(\beta)$, which would result in an overly anti-symmetric current density distribution. $I_{c,\text{even}}(\beta)$ is found by multiplying the translated $I_c$ by a flipping function that changes the sign of each subsequent lobe of the interference pattern, as can be observed in Figure \ref{FIGS2}b. We follow the above procedure for finding $I_{c,\text{odd}}(\beta)$, which is depicted in Figure \ref{FIGS2}c. The corresponding critical current density distribution is found by a numerical Fourier transform carried out in Python using the Numpy package, yielding the distribution in Figure \ref{FIGS2}d.

The original work of Dynes and Fulton describes an overlap junction between two superconductors that are relatively thick (\textit{i.e.}, the superconductors extend over multiple times $\lambda$ on each side of the junction. In this case, both superconductors can effectively shield out the magnetic field, yielding an effective junction length $L = 2 \lambda + d$, where $d$ is the barrier thickness. The disk devices, however, are planar junctions with a film thickness that is small in comparison to the London penetration depth ($\lambda(0) = 140$ nm). In this limit, the sine-Gordon equation is modified due to non-local electrodynamics and flux focusing effects, and the period of the SQI pattern is only determined by the geometry of the junction.\cite{Abdumalikov2009,Moshe2008,Boris2013} Generally, in this limit $L_{\text{eff}} = 0.56W$ is found. As expected, for our devices, the period of the SQI oscillations scales with the radius of the disk. We find for both radii that:

\begin{align} \label{EqS5}
\frac{\Phi_0}{\Delta B (2R)^2} = \frac{\Phi_0}{\Delta B W^2} = 0.26
\end{align}

\noindent where $\Delta B$ is the field spacing of the peaks in the SQI pattern. The effective length is then determined as:

\begin{align} \label{EqS6}
L_{\text{eff}} = \frac{A_{\text{eff}}}{W} = \frac{\Phi_0}{\Delta B W} = 0.26W
\end{align}

\begin{figure*}[bt!]
 \centerline{$
 \begin{array}{c}
  \includegraphics[width=0.5\linewidth]{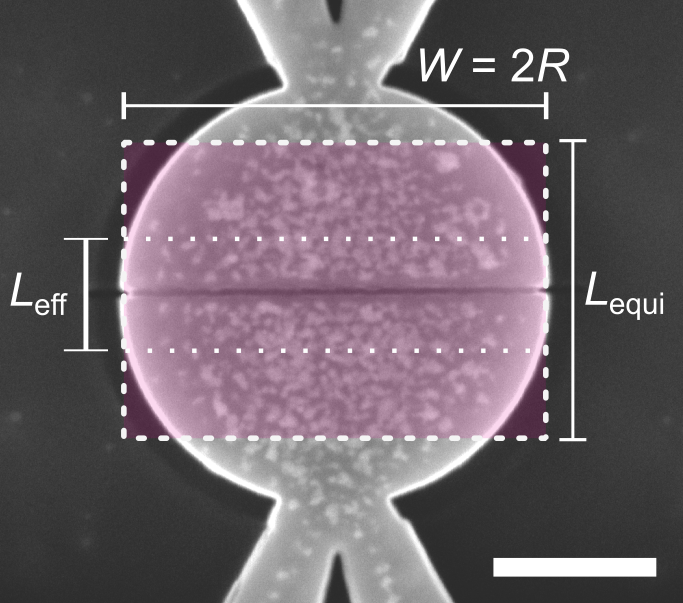}
 \end{array}$}
 \caption{SEM micrograph of a disk-device ($R$ = 1.7 $\mu$m), depicted with the length scales relevant to the Fourier analysis. The shaded region corresponds to an equivalent rectangular planar junction, considered in \cite{Clem2010}, where $L_{\text{equi}} = 0.7W$. $L_{\text{eff}} = 0.26W$ holds both for our disk devices and the equivalent rectangular junction. The scale bar equals 500 nm.} \label{FIGS_extra}
\end{figure*}

\noindent Resulting in a prefactor of $L_{\text{eff}}$ that is significantly lower for our disk junctions. However, if the shielding currents are also restricted in the lateral dimensions, it is shown that the prefactor can decrease.\cite{Clem2010} Translated to an equivalent rectangular geometry described in \cite{Clem2010}, $L_{\text{eff}} = 0.26W$ corresponds to a junction with a total length of $L_{\text{equi}} = 0.7W$. A comparison between such geometry and our disk devices is shown in Figure \ref{FIGS_extra}. Evidently, $L_{\text{equi}}$ is close to the diameter of the disk, giving a strong indication for the validity of the use of $L_{\text{eff}} = 0.26W$ in the Fourier analysis. The fact that $L_{\text{equi}}$ is slightly lower than the diameter can be explained by the circular shape of the sample boundaries, which causes a further restriction to the shielding currents.

It is important to note that the spatial resolution of current density distribution is determined by the number of lobes in the SQI pattern, which is limited by the possible discontinuities in the SQI pattern (typical to junctions with a ferromagnetic weak link). This sets a bound to the precision in determining the width of the current channels. For all measured disk junctions we find the locations of the current channels within 70 nm (binning size of the Fourier transform) from the edges.

Finally, strictly speaking, part of the rectangular junction area that is used in the Fourier analysis falls outside the disk geometry. Using the above-mentioned values of $L_{\text{eff}}$, this concerns a mere 6\% of the total area used in the calculation. Therefore, we can assume the analysis remains valid up to a high degree for our devices.

\section{Details of theoretical calculations}

\subsection{Basic equations}

To analyze a proximity effect in ferromagnet/superconductor system we use
linearised Usadel equation for the anomalous function which takes
into account triplet and singlet superconducting
correlations. We presented anomalous function in the
ferromagnetic region in the form 
\begin{align}
\hat{f}=f_s\hat{\sigma}_0+ i (\bm f \bm \sigma)
\end{align}
 In this expansion the first term corresponds to the singlet
component and the last three terms correspond to the triplet
components of the anomalous function. 
The equations for components where we include the spin-flip scattering are (for $\omega>0$):
 \begin{align}\label{Eq:UsadelSpinFLip1}
 & [D\nabla^2-2(\omega + 3/\tau_{sf}) ]f_s 
 + 
 2 ({\bm h} {\bm f})=0,
 \\ \label{Eq:UsadelSpinFLip2}
 & [ D\nabla^2 - 2(\omega + 1/\tau_{sf}) ] {\bm f} 
 - 
  2f_s {\bm h} =0 
 \end{align}  
 with the boundary conditions
 \begin{align}\label{Eq:bcSinglet}
  & ({\bm n} \nabla) f_s |_F = \gamma \hat f_s|_S  
  \\ \label{Eq:bcTriplet}
  & ({\bm n} \nabla) {\bm f} |_F = 
  \gamma {\bm f} |_S  
 \end{align}
 where $\gamma$ is the transparency of the F/S interface,
 $D$ is diffusion coefficient, ${\bm h}$  is the exchange
 field proportional to the magnetic moment and $\omega$ is the
 Matsubara frequency. We add the spin relaxation term with rate $\tau_{sf}^{-1}$.
 The sources in the right-hand side of 
 Eq. \ref{Eq:bcSinglet} and \ref{Eq:bcTriplet}
  $\hat f_s|_S $ and ${\bm f} |_S $
 are the spin-singlet and spin-triplet   
  components of the anomalous function in the superconducting electrode.
 In the simplest case it contains only the spin-singlet part 
  $f_s|_S=\hat \Delta/\sqrt{|\Delta|^2+\omega^2}$ and ${\bm f} |_S =0$. 
  However 
  in general there are also the spin-triplet correlations 
  generated due to the inverse proximity effect
  \cite{bergeret2004induced,bergeret2005inverse}
  or due to the magnetic proximity effect\cite{Tokuyasu1988,cottet2009spin,Eschrig2015} through the spin-mixing mechanism.

   If the distance between superconducting electrodes is large the 
   current though F is mediated only by the long-range spin-triplet (LRT) components  with $\bm f_{LRT} \perp \bm m$, where $\bm m$ is the local magnetization direction.

 \subsection{Generation of LRT from spin supercurrent conservation}
 
 \subsubsection{Qualitative picture}
 
Let us consider the overlap geometry when both the S/F interface and the inhomogeneity of the magnetization gradients are in the  $xy$ plane. 
 For definitiveness let us assume that magnetization rotates in $xy$-plane
$\bm m =  ( -\sin\theta_v, \cos\theta_v,0)$. 
For example, in the magnetic vortex 
$   \theta_v = \arctan [(y-y_0)/(x-x_0)]
$.
It is convenient to transform the magnetic texture into the homogeneous one by the spin rotation 
\begin{align}
   \hat f \to  
   e^{i (\theta_v-\pi/4) \hat\sigma_z }  
    \hat f 
    e^{-i (\theta_v-\pi/4)\hat\sigma_z }   ;  \; 
   \end{align}
 This transformation makes magnetic texture uniform with 
 $ m_x = 1$ and $m_z=m_y=0$. In addition, it leads to the effective spin-orbital coupling (SOC)
  which shows up in the expression for the elongated spatial derivative 
   \begin{align}
    \label{Eq:DCovariant}
    \nabla_j \hat f \to  
    \nabla_j \hat f + 
   i   Z_j [\hat\sigma_z, \hat f] 
   = 
   \hat{\bm\sigma} 
   ( \nabla_j \bm f  -  Z_j [\bm z\times \bm f] )  
   \end{align} 
   with the spin-dependent vector potential 
   $\bm Z  = \nabla \theta_v $. The term $\hat\sigma_z \bm Z$ plays the role of effective spin-orbital coupling. 
   
 \begin{figure*}[t!]
 \centerline{$
 \begin{array}{c}
  \includegraphics[width=1\linewidth]{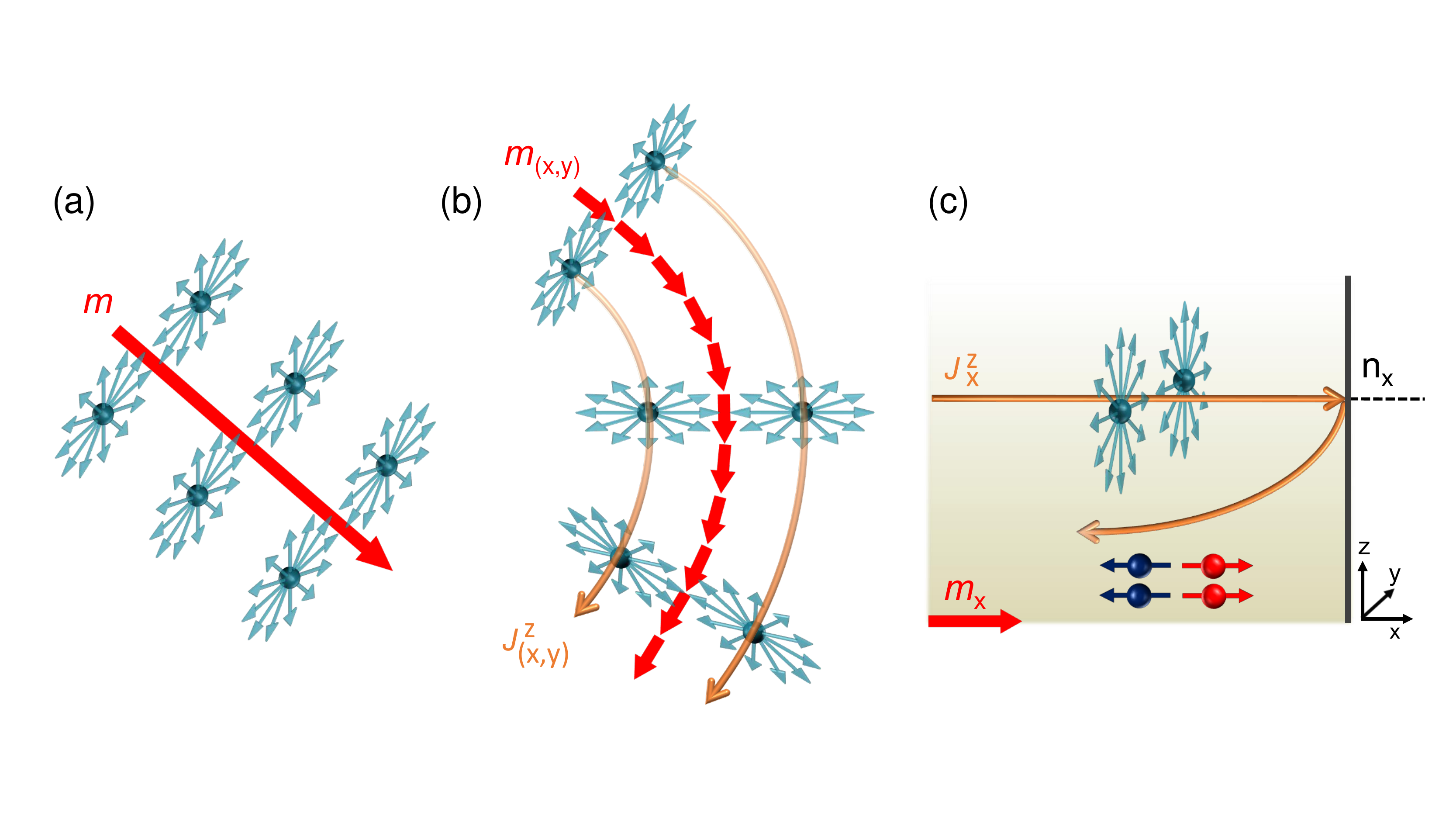}
 \end{array}$}
 \caption{\label{Fig:6} \textbf{a.} SRT Cooper pairs (cyan) in a system with uniform magnetization. The SRT correlations have zero spin projection along the quantization axis, defined by the local magnetization vector $\bm m$ (red arrow). No equilibrium spin current (ESC) or LRT pairs are generated. 
 \textbf{b.} In the presence of spin texture, the SRT pairs continuously alter the orientation of their spin plane to remain normal to the local magnetization vector $\bm m_{(x,y)}$, producing an ESC $\bm J_{(x,y)}^z$ (orange arrows), which adiabatically follows the local spin gauge field $\nabla\theta_v = (\bm m\times \nabla \bm m)_z $. \textbf{c.} When the ESC is incident on a bilayer-vacuum boundary, the adiabatic approximation breaks down. As the ESC cannot propagate outside the bilayer, the total spin current at the boundary must remain zero. To compensate for the adiabatic spin current, a condensate of LRT pairs (in red and blue) with $S_x=\pm 1$ emerges near the vacuum boundary to generate an opposing spin current.
} \label{Co-fig6}
 \end{figure*}   

The transformation (\ref{Eq:DCovariant})
 demonstrates the importance of edges in generating the 
 spin-triplet components which are not collinear with magnetization and therefore can penetrate into the ferromagnet.
 Indeed, far from the magnetic vortex core, the scale of changing magnetization texture is large, that the emergent vector potential field can be considered as locally-homogeneous 
 $\bm Z=const$
 which corresponds to the magnetic texture changing in 
 space with constant gradient.
 In this case, far from the edges only the correlations 
 with $\bm f \parallel \bm h$ exist in S and in 
 F in the thin layer of the size $\xi_F= \sqrt{D/h}$ near the S/F interface. 
     However even the constant spin-dependent 
   gauge field $\bm Z$ generates the equilibrium spin current 
   \cite{Alidoust2015_discus1, Alidoust2015_discus2, Tokatly2019}
    \begin{align} 
  \label{Eq:SpinCurrent1}
 \bm J_j = 
 \frac{\pi\sigma_n T }{e^2}
  \sum_\omega  
 ( \bm f \times \nabla_j \bm f - 
 Z_j |\bm f|^2 \bm z )
 \end{align}     
  
  At the superconductor/vacuum or ferromagnet/vacuum boundaries the spin current should vanish 
  $ n_j \bm J_j =0$.
 Using (\ref{Eq:SpinCurrent1}) this boundary condition can be written as 
  \begin{align} \label{Eq:BCcorrection}
  & (\bm n  \nabla) \bm f = - (\bm n\bm Z) [\bm z\times \bm f]
    \end{align}  
    It provides the coupling between different spin components of the anomalous function.
 That is the r.h.s. provides 
  sources of $\bm f \perp \bm m$.   
  Indeed since in the local spin-rotated frame 
    $\bm m \parallel \bm x$  only ${\bm f}= f_x\bm x$ exist far from the edge. But substituting it to the r.h.s. of (\ref{Eq:BCcorrection}) we get the sources of the 
     transverse component 
  $\bm f_{LRT} = f_y \bm y $. 
      Note that this boundary condition (\ref{Eq:BCcorrection}) is valid both in S and F. 
      
      The generation of spin-triplet correlations with $\bm f_{LRT} \perp \bm m$ leads to the edge accumulation of the transverse spin component \cite{Alidoust2015_discus1,Alidoust2015_discus2, Tokatly2019}. 
 Such correlations can propagate in the ferromagnet at the distances determined by the normal coherence length $\xi_N = \sqrt{D/T}$ or the spin-flip relaxation length $l_s = \sqrt{D\tau_{sf}} $. Therefore we call such correlations as long-range triplet correlations (LRT).

  
  The boundary condition (\ref{Eq:BCcorrection}) provides the generation of LRT both in the S and F electrodes. 
  The generation in F electrode is described by the linearized Usadel 
  Eqs. (\ref{Eq:UsadelSpinFLip1},\ref{Eq:UsadelSpinFLip2}) 
  and boundary conditions 
  (\ref{Eq:bcSinglet}, \ref{Eq:bcTriplet}) with r.h.s. 
  $f_s |_S= F_{0}=\Delta /\sqrt{\Delta^2 + \omega^2}$ 
 and $\bm f|_S=0$. 
  
 The generation in S electrode is in general determined by the 
 non-linear Usadel equations with inhomogeneous Zeeman field $\bm h\parallel \bm m $ which can be induced either by the inverse proximity effect or the 
 magnetic proximity effect. 
 In the case of inhomogeneous $\bm h$ we again obtain a non-zero LRT source 
 $\bm f|_S \perp \bm h$ in the r.h.s. of the Eq. \ref{Eq:bcTriplet}. 
    In both scenarios, the generation of LRT in the presence of a magnetic vortex is described by the qualitatively similar equations derived below; a schematic of the LRT generation is shown in Figure \ref{Co-fig6}.

  \subsubsection{Generation of LRT in S and F electrodes}
 
 In general the superconducting  correlations both in S and in F are determined by the Usadel equation for the quasiclassical Green function (GF)
 \begin{align} \label{Eq:UsadelGenS}
  \nabla (D\hat g \nabla   \hat g)  
  - 
 [\hat\Delta+\omega\hat \tau_3 + 
 i  \bm h \bm \sigma \hat\tau_3, \hat g] =0 
  \end{align}
 Here $\hat\tau_{1,2,3}$ are the Pauli matrices in Nambu space.    
   The GF has the off-diagonal anomalous part $\hat \tau_1  \hat f$ which describes the 
   superconducting correlations.
       The field $\bm h$  can be either
          {\bf (i)} the induced Zeeman field in S
      or {\bf (i)} the  real exchange field  inside F. 
    We consider these two cases separately. 
    
   {\bf (i)} First, let us assume that $\bm h$ is the Zeeman field in S, homogeneous over the $z$ coordinate. 
    The local approximation produces the collinear triplet     $f_{SRT} \bm m$ which in the simplest case is given by  
    $
 f_{SRT} = [ f_0 (\omega + i h) -
  f_0 (\omega - i h) ]/2
  $ 
  where  $f_0$ is the anomalous function in the absence of the Zeeman field. The first-order correction by $\nabla \bm m$ has the form of the transverse   spin component $\bm f_{LRT} \perp \bm  m$
  and appears only because of the boundary condition $\bm n\nabla \hat g=0$ near the  edges.  
  The equation is simplified in case that the coplanar magnetization
 texture $m_z=0$. Then the transverse spin vector can be found in the form $\bm f_{LRT} = (-m_y, m_x, 0) \Psi_{LRT} $.
      We assume that the transverse spin-triplet component is small and linearise (\ref{Eq:UsadelGenS}) with respect to $\Psi_{LRT}$.
 The amplitude satisfy equations
 \begin{align} \label{Eq:UsadelLRSuper}
 & \nabla^2_\perp   
 \Psi_{LRT} - \xi_\omega^{-2}  \Psi_{LRT} 
 = 
 0
 \\ \label{Eq:UsadelLRbcSuper}
 & (\bm n \nabla) \Psi_{LRT} = 
 -  
  \bm n (\bm m \times \nabla \bm m)_z 
   \Psi_{SRT}
   \end{align}       
 where $\nabla_\perp = (\nabla_x, \nabla_y, 0)$, 
  $\xi^{-2}_\omega = 2(\omega^2 + |\Delta|^2)/D$.
 The equation is simplified in case of the coplanar magnetization
 texture $m_z=0$. Then the transverse spin vector can be found in the form $\bm f_{LRT} = (-m_y, m_x, 0) \Psi_{LRT} $. Solving Eqs.(\ref{Eq:UsadelLRSuper},\ref{Eq:UsadelLRbcSuper}) we find the distribution of $\Psi_{LRT}$ in the superconductor. 
Then, using the boundary conditions (\ref{Eq:bcTriplet})
 we obtain the distribution of LRT in the F  from the equations 
 \begin{align} \label{Eq:bcLRTF}
 & \nabla^2
 \Psi_{LRT} - \lambda_\omega^{-2}   \Psi_{LRT}  =  0
  \\  
 \label{Eq:bcLRTzF}
 & \partial_z \Psi_{LRT}|_F = \gamma \Psi_{LRT} |_S  
 \end{align}
  Note that the Eqs.(\ref{Eq:bcLRTF}) is similar to (\ref{Eq:UsadelLRSuper}) with the only difference in the decay 
  scale determined by $\lambda_\omega = 
  1/\sqrt{\xi_\omega^{-2} + \lambda_{sf}^{-2}}$ where $\xi_\omega = \sqrt{D/|\omega|}$ and $\lambda_{sf} $ is the normal metal spin diffusion length. 
  
  In case if the F thickness is smaller than the normal metal coherence length and spin diffusion length we can integrate 
  Eqs.(\ref{Eq:bcLRTF},\ref{Eq:bcLRTzF}) by $z$ to get the equation with the source 
 \begin{align} \label{Eq:bcLRTFSource}
 & \nabla^2_\perp  \Psi_{LRT} - \lambda_\omega^{-2} \Psi_{LRT} 
 =  (\gamma/d_F) \Psi_{LRT} |_S 
   \end{align}
   Solving this equation we obtain the distributions of LRT show in Figure 5 of the main text.  
 
 \begin{figure}[t]
 \centerline{$
 \begin{array}{c}
 \includegraphics[width=0.40\linewidth]{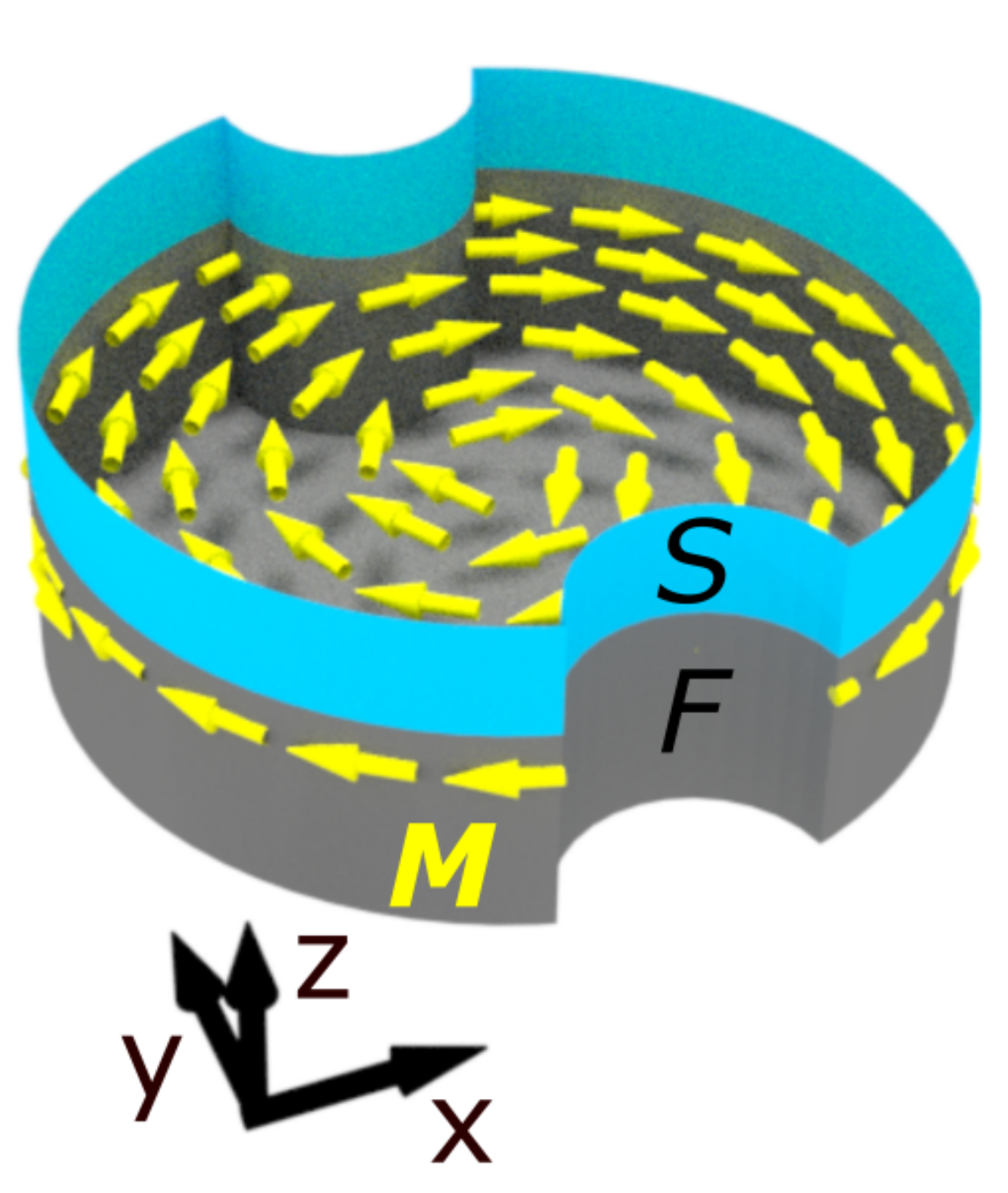} \;\;
    \end{array}$}
 \caption{\label{Fig:1-1} 
  Sketch of the overlap 
 S/F structure with magnetic vortex pattern in the lower magnetic layer. The notches can be seen on the side.
    }
 \end{figure}   
  
 {\bf (ii)}
  The SRT can be generated by the exchange field in F
  determined by the linearized Usadel Eqs.(\ref{Eq:UsadelSpinFLip1},\ref{Eq:UsadelSpinFLip2}). 
 For large exchange field $h \gg T_c$ they have 
    solutions
of two types which are the short-range and long-range modes with
the scales $\xi_F=\sqrt{D/h}$ and $\xi_\omega=\sqrt{D/|\omega|}$
correspondingly. Hereafter we assume that $\xi_F$ is the smallest
length of the problem such that the spatial dependencies of
exchange field and geometrical factors are characterized by the
scales $\gg \xi_F$. 

The spin-triplet correlations can be written as the superposition of short and long-range modes
 $\bm f= \bm f_{SRT} + \bm f_{LRT}$. 
 The spin vector of short range modes is parallel to the 
 exchange field 
 $\bm f_{SRT} = \bm m \Psi_{SRT}$.  
Under such conditions we search the
short-range solutions of Eqs.(\ref{Eq:UsadelSpinFLip1},\ref{Eq:UsadelSpinFLip2}) in the form
 \begin{align} \label{Eq:SRSC0}
 & f_s (\bm r) = - i\gamma
\xi_F F_{bcs} 
 ( e^{i\pi/4} e^{- \lambda_h z} - c.c.) /2
 \\  \label{Eq:SRTCz}
 & \Psi_{SRT} = - \gamma
\xi_F F_{bcs}  
 ( e^{i\pi/4} e^{- \lambda_h z} + c.c.) /2
 \end{align}
 where the scale is $\lambda_h =  e^{i\pi/4}/\xi_F$. In case if the thickness of F is smaller than the coherence and spin diffusion lengths  $d_F$ the LRT distribution is determined by the equation similar to Eq. \ref{Eq:UsadelLRSuper}
 \begin{align} \label{Eq:UsadelLRF}
 & \nabla^2_\perp  \Psi_{LRT} - \lambda_\omega^{-2} \Psi_{LRT} 
 =  0 
 \\ \label{Eq:UsadelLRbcF}
 & (\bm n \nabla) \Psi_{LRT} = 
 - \bm n (\bm m \times \nabla \bm m)_z 
  \langle\Psi_{SRT}\rangle_z
   \end{align}
   where the $z$-averaged SRT is 
   $ \langle\Psi_{SRT}\rangle_z = (\gamma /d_F) (T_{c}/ h)  F_{bcs}$.

  \subsubsection{Example: LRT generated by the magnetic vortex}

 The distribution of magnetization in the magnetic vortex far from the vortex core is given by 
  $\bm m =  (- sin\theta_v, \cos\theta_v, 0) $ where $\theta_v$ is the polar angle with respect to the vortex center. 
  In addition, we  take into account the non-circular geometric shape, \textit{e.g.}, shown in Figure \ref{Fig:1-1}. 
We  find the LRT distribution solving either the system 
(\ref{Eq:UsadelLRSuper}, \ref{Eq:UsadelLRbcSuper},\ref{Eq:bcLRTFSource})
or (\ref{Eq:UsadelLRF}, \ref{Eq:UsadelLRbcF}) using finite element numerical package FreeFem\cite{MR3043640}. 
Both systems yield qualitatively similar LRT distributions shown in Figure 5 of the main text.

 \begin{figure*}[t!]
 \centerline{$
 \begin{array}{c} 
       \includegraphics[width=1.0\linewidth]{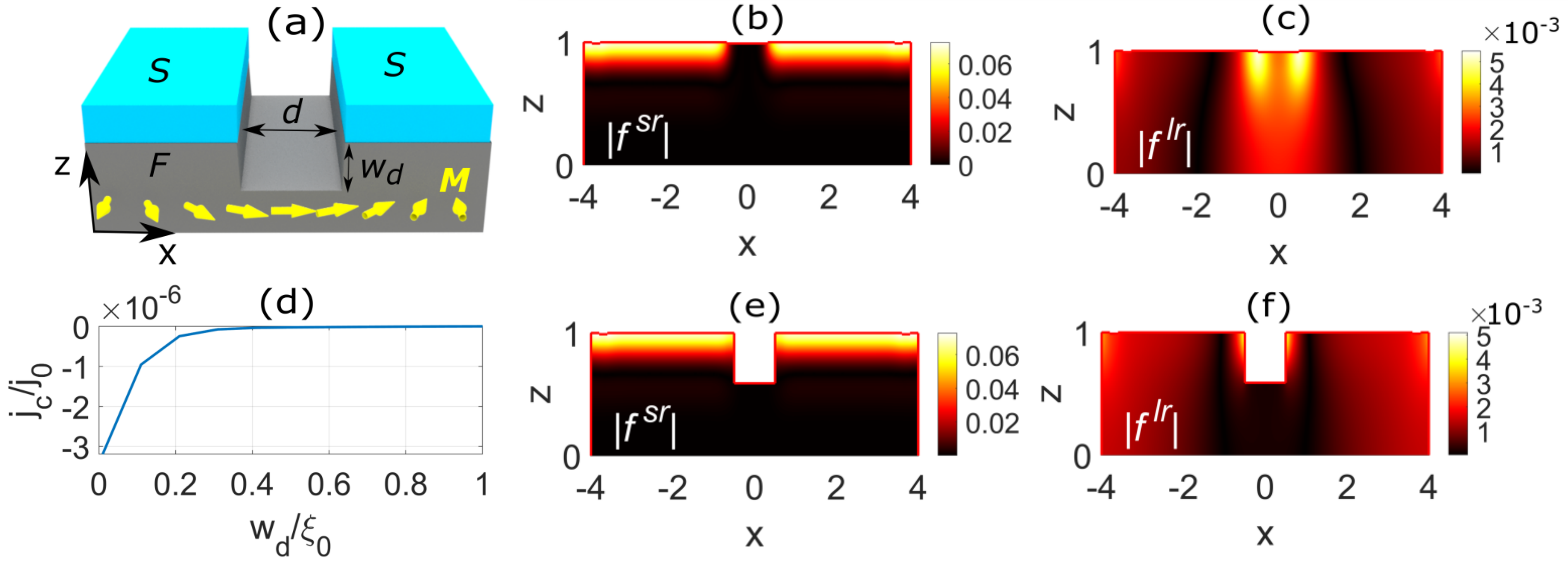}
           \end{array}$}
 \caption{\label{Fig:3} 
 Spin-triplet superconducting correlations 
 in the model overlap S/F/S junction. 
  (a) Sketch of the setup. The magnetization pattern 
is $\bm M=M_0 \bm m$ where $\bm m = ( a , x, 0)/\sqrt{x^2 + a^2}$ and $a = 2\xi_0$. 
 Exchange field is $h=100 T_{c}$, temperature $T=0.1 T_c$ and spin relaxation rate $\tau_{sf} T_{c}=0.25$, the distance between S electrodes is $d=\xi_0$.
The phases in $S_{1,2}$ 
 are $\varphi_{1}- \varphi_2=\pi$. The geometric defect in F of the depth $w_d$  is in the trench. 
 (b,c) Distributions SRT. (c,f) Distributions of LRT.   
    (d) Critical current as a function of the depth of the trench $w_d$. 
       }
 \end{figure*}   

\subsection{Generation of triplets at the trench}
Besides the boundary sources of LRTs, some sources come directly from the Usadel equation in the presence of non-magnetic inhomogeneities such as the order parameter profile.\cite{Kalenkov2011} Indeed, the inhomogeneous  $|\bm f_{SRT}|^2$ yields the  divergence of the second term in the spin current  (\ref{Eq:SpinCurrent1}) $\nabla_j\bm J_j \propto Z_j \nabla_j |\bm f_{srt}|^2 \bm z$. Such sources have to be compensated by the contribution from the LRT.\\
\indent To support the above qualitative arguments we consider the generic 2D problem of a superconductor/ferromagnet bilayer shown in Figure \ref{Fig:3}. We assume that exchange field amplitude is $h=100 T_{c0}$ and  the magnetization pattern is $\bm M = M( a , x, 0)/\sqrt{x^2 + a^2}$ with $a = 2\xi_0$. The F-layer has a thickness of $\xi_0$ and is $8\xi_0$ wide, where $\xi_0= \sqrt{D/T_{c}}$. \\
\indent We consider two model overlap S/F/S junctions shown in Figure \ref{Fig:3}. The upper row corresponds to smooth edges of S electrodes parametrized by the coordinate-dependent transparency of the S/F interface. The lower row corresponds to the sharp edges of S electrode with $\gamma (x)$ changing abruptly. By solving the linearised Usadel equations (\ref{Eq:UsadelSpinFLip1}, \ref{Eq:UsadelSpinFLip2}, \ref{Eq:bcSinglet}, \ref{Eq:bcTriplet}) with $f_s|_S=\hat \Delta/\sqrt{|\Delta|^2+\omega^2}$ and ${\bm f} |_S =0$, using the finite element numerical package FreeFem,\cite{MR3043640} we can separate the SRT contribution $\bm f_{SRT} = f_s \bm h/h$ and the LRT one $\bm f_{LRT} = \bm f  - \bm f_{SRT}$ for several different configurations. This allows us to study formation of LRT correlations near the edges of F and S layers. \\
\indent The fourth row showing the total LRT amplitude $|f_{LRT}|$, indicates that they are indeed generated  at the boundaries of F layer. Also, LRT correlations are generated  near the edges of the overlap S layers which forms the trench. At the same time, this source of LRT generation is very sensitive to the presence of surface defects at the trench. The destructive influence of the defect can be seen in Figure \ref{Fig:3} comparing the panels (c) and (f), with a flat surface and the rectangular well between the S electrodes. The suppression of LRT generation results in the strong suppression of the critical current $j_c$ shown in panel (d). The current is normalized to $j_0 = \sigma_n T_c /e  $ which is of the order of $10^2/\xi_0$  A/cm. Such current flowing through the cross section of F of the linear size $R\sim 1 \; \mu $m yields $j\sim 10^{-8} $ A, which is three orders of magnitude smaller than the experimentally observed value.
\clearpage


%